\documentclass[10pt,journal,compsoc]{IEEEtran}

\usepackage{multirow}
\usepackage{array}
\usepackage{subfigure}
\usepackage{amsmath}
\usepackage{graphicx}
\usepackage{epstopdf}
\usepackage{amsfonts}
\usepackage{url}
\usepackage{booktabs}

\ifCLASSOPTIONcompsoc
  \usepackage[nocompress]{cite}
\else
  \usepackage{cite}
\fi

\ifCLASSINFOpdf
\else
\fi
\begin{document}
%
\title{Neural Diffusion Model for\\ Microscopic Cascade Prediction}
%
%
%
%

\author{Cheng Yang, Maosong Sun, Haoran Liu, Shiyi Han, Zhiyuan Liu, Huanbo Luan
\IEEEcompsocitemizethanks{\IEEEcompsocthanksitem Cheng Yang, Maosong Sun (corresponding author), Zhiyuan Liu and Huanbo Luan are with the Department of Computer Science and Technology, Tsinghua University, Beijing 100084, China.\protect\\
E-mail: cheng-ya14@mails.tsinghua.edu.cn, \protect\\\{sms,liuzy\}@mail.tsinghua.edu.cn, luanhuanbo@gmail.com
\IEEEcompsocthanksitem Haoran Liu is with the Department of Electric Engineering, Tsinghua University, Beijing 100084, China.\protect\\
E-mail: liu-hr15@mails.tsinghua.edu.cn
\IEEEcompsocthanksitem Shiyi Han is with the Department of Computer Science, Brown University, U.S.A.\protect\\
E-mail: hanshiyi123@gmail.com}}

%
%

\markboth{Transactions on Knowledge and Data Engineering,~Vol.~14, No.~8, August~2015}%
{Yang \MakeLowercase{\textit{et al.}}: Neural Diffusion Model for Microscopic Cascade Prediction}

\IEEEtitleabstractindextext{%
\begin{abstract}
The prediction of information diffusion or cascade has attracted much attention over the last decade. Most cascade prediction works target on predicting cascade-level \textit{macroscopic} properties such as the final size of a cascade. Existing \textit{microscopic} cascade prediction models which focus on user-level modeling either make strong assumptions on how a user gets infected by a cascade or limit themselves to a specific scenario where ``who infected whom'' information is explicitly labeled. The strong assumptions oversimplify the complex diffusion mechanism and prevent these models from better fitting real-world cascade data. Also, the methods which focus on specific scenarios cannot be generalized to a general setting where the diffusion graph is unobserved.

To overcome the drawbacks of previous works, we propose a Neural Diffusion Model (NDM) for general microscopic cascade prediction. NDM makes relaxed assumptions and employs deep learning techniques including attention mechanism and convolutional network for cascade modeling. Both advantages enable our model to go beyond the limitations of previous methods, better fit the diffusion data and generalize to unseen cascades. Experimental results on diffusion prediction task over four realistic cascade datasets show that our model can achieve a relative improvement up to $26\%$ against the best performing baseline in terms of F1 score.
\end{abstract}

\begin{IEEEkeywords}
Information Diffusion, Neural Network
\end{IEEEkeywords}}

\maketitle

\IEEEdisplaynontitleabstractindextext

%
\IEEEpeerreviewmaketitle

\IEEEraisesectionheading{\section{Introduction}\label{sec:introduction}}

\IEEEPARstart{I}nformation diffusion is a ubiquitous and fundamental event in our daily lives, such as the spread of rumors, the contagion of viruses and the propagation of new ideas and technologies. The diffusion process, also called a \textit{cascade}, has been studied over a broad range of domains. Though some works believe that even the eventual size of a cascade cannot be predicted~\cite{salganik2006experimental}, recent works~\cite{cheng2014can,yu2015micro,zhao2015seismic} have shown the ability to predict the size, growth and many other key properties of a cascade. Nowadays the modeling and prediction of cascades play an important role in many real-world applications, \textit{e.g.} production recommendation~\cite{domingos2001mining,leskovec2006patterns,leskovec2007dynamics,watts2007influentials,aral2012identifying}, epidemiology~\cite{hethcote2000mathematics,wallinga2004different}, social networks~\cite{kempe2003maximizing,lappas2010finding,dow2013anatomy} and the spread of news and opinions~\cite{gruhl2004information,liben2008tracing,leskovec2009meme}. 

\begin{figure}[tb]
\centering
\includegraphics[width=.99\columnwidth]{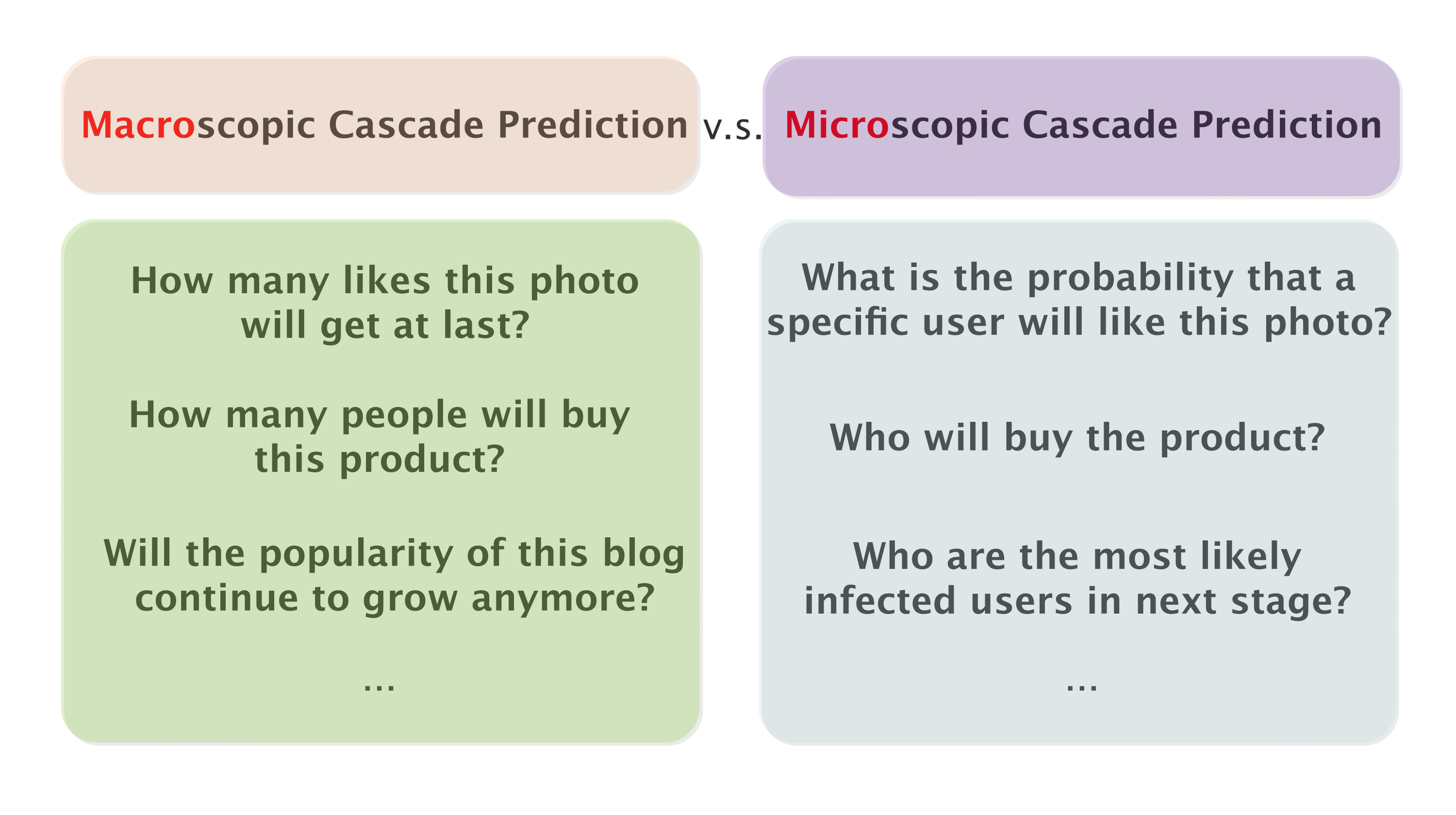}
\caption{Macroscopic cascade prediction v.s. microscopic cascade prediction.}
\label{fig:intro}
\end{figure}

Most previous works on cascade prediction focus on the prediction of \textit{macroscopic} properties such as the total number of users who share a specific photo~\cite{cheng2014can} and the growth curve of the popularity of a blog~\cite{yu2015micro}. However, macroscopic cascade prediction is a rough estimate of cascades and cannot be adapted for \textit{microscopic} questions as shown in Fig.~\ref{fig:intro}. Microscopic cascade prediction, which pays more attention to user-level modeling and prediction instead of cascade-level, is much more powerful than macroscopic prediction and allows us to apply user-specific strategies for real-world applications. For example, during the adoption of a new product, microscopic cascade prediction can help us deliver advertisements to those users that are most likely to buy the product at each stage. In this paper, we focus on the study of microscopic cascade prediction.

Though useful and powerful, the microscopic prediction of cascades faces great challenges because the real-world diffusion process could be rather complex~\cite{romero2011differences} and usually partially observed~\cite{wallinga2004different,myers2010convexity}:

\textbf{Complex mechanism.} Since the mechanism of how a specific user gets infected~\footnote{We use ``infected'' and ``activated'' alternatively to indicate that a user is influenced by a cascade.} is sophisticated, traditional cascade models based on strong assumptions and simple formulas may not be the best choice for microscopic cascade prediction. Existing cascade models~\cite{gomez2010inferring,rodriguez2014uncovering,gomez2013structure,bourigault2016representation} which could be adopted for microscopic prediction mostly ground in Independent Cascade (IC) model~\cite{kempe2003maximizing}. IC model assigns a static probability $p_{u,v}$ to user pairs $(u,v)$ with pairwise independent assumptions, where the probability $p_{u,v}$ indicates how likely user $v$ will get infected by user $u$ when $u$ is infected. Other diffusion models~\cite{bourigault2014learning,gao2017novel} make even stronger assumptions that the infected users are only determined by the source user. Though intuitive and easy to understand, these cascade models are based on strong assumptions and oversimplified probability estimation formulas, both of which limit the expressivity and ability to fit complex real-world cascade data~\cite{li2017deepcas}. The complex mechanism of real-world diffusions encourages us to explore more sophisticated models, \textit{e.g.} deep learning techniques, for cascade modeling.

\textbf{Incomplete observation.} On the other hand, the cascade data is usually partially observed indicates that we can only observe those users getting infected without knowing who infected them. However, to the best of our knowledge, existing deep-learning engined microscopic cascade models~\cite{hu2017will,wang2017topological} are based on the assumption that the diffusion graph where a user can only infect and get infected by its neighbors is already known. For example, when we study the retweeting behavior on the Twitter network, ``who infected whom'' information is explicitly labeled in retweet chain and the next infected user candidates are restricted to the neighboring users rather than the whole user set. While in most diffusion processes such as the adoption of a product or the contamination of a virus, the diffusion graph is unobserved~\cite{wallinga2004different,myers2010convexity,kefato2017di}. Therefore, these methods consider a much simpler problem and cannot be generalized to a general setting where the diffusion graph is unknown.

To fill in the blank of general microscopic cascade prediction and address the limitations of traditional cascade models, we propose a neural diffusion model based on relaxed assumptions and employ up-to-date deep learning techniques, \textit{i.e.} attention mechanism and convolutional neural network, for cascade modeling. The relaxed assumptions enable our model to be more flexible and less constrained, and deep learning tools are good at capturing the complex and intrinsic relationships that are hard to be characterized by hand-crafted features. Both advantages allow our model to go beyond the limitations of traditional methods based on strong assumptions and oversimplified formulas and better fit the complicated cascade data. Following the experimental settings in~\cite{bourigault2016representation}, we conduct experiments on diffusion prediction task over four realistic cascade datasets to evaluate the performances of our proposed model and other state-of-the-art baseline methods. Experimental results show that our model can achieve a relative improvement up to $26\%$ against the best performing baseline in terms of F1 score.

To conclude, our contributions are $3$-fold:
\begin{itemize}
\item To the best of our knowledge, our work is the first attempt to employ deep learning techniques for general microscopic cascade prediction problem where the diffusion graph is unknown.
\item We design a neural diffusion model based on relaxed assumptions compared with the pairwise independence assumption in traditional cascade models and allow our model to better fit real-world cascades and generalize to unseen data.
\item Experimental results on diffusion prediction task over four realistic datasets demonstrate the effectiveness and robustness of our proposed model. Compared with the best performing baseline, our model can achieve a relative improvement up to $26\%$ on F1 score.
\end{itemize}

\section{Related Works}
We organize related works into macroscopic and microscopic cascade prediction methods. In terms of methodology, our work is also related to network representation learning methods.
\subsection{Macroscopic Cascade Prediction}
Most previous works on cascade prediction focused on macroscopic level prediction such as the eventual size of a cascade~\cite{zhao2015seismic} and the growth curve of popularity~\cite{yu2015micro}. Macroscopic cascade prediction methods can further be classified into feature-based approaches, generative approaches, and deep-learning based approaches. Feature-based approaches formalized the task as a classification problem~\cite{cui2013cascading,cheng2014can} or a regression problem~\cite{tsur2012s,weng2014predicting} by applying SVM, logistic regression and other machine learning algorithms on hand-crafted features including temporal~\cite{pinto2013using} and structural~\cite{cheng2014can} features. Generative approaches considered the growth of cascade size as an arrival process of infected users and employed stochastic processes, such as Hawkes self-exciting point process~\cite{gao2015modeling,zhao2015seismic}, for modeling. With the success of deep learning techniques in various applications, deep-learning based approaches, e.g. DeepCas~\cite{li2017deepcas} and DeepHawkes~\cite{cao2017deephawkes}, were proposed to employ Recurrent Neural Network (RNN) for encoding cascade sequences into feature vectors instead of hand-crafted features. Compared with hand-crafted feature engineering, deep-learning based approaches have better generalization ability across different platforms and give better performance on macroscopic prediction task.

\subsection{Microscopic Cascade Prediction}
Our work is more related to microscopic cascade prediction which focuses on user-level modeling and predictions. We classify related works into three categories: IC-based approaches, embedding-based approaches, and deep-learning based approaches.

IC model~\cite{goldenberg2001talk,kempe2003maximizing,gruhl2004information,saito2008prediction} is one of the most popular diffusion models which assumed independent diffusion probability through each link. Extensions of IC model further considered time delay information by incorporating a predefined time-decay weighting function, such as continuous time IC~\cite{saito2009learning}, CONNIE~\cite{myers2010convexity}, NetInf~\cite{gomez2010inferring} and Netrate~\cite{rodriguez2014uncovering}. Infopath~\cite{gomez2013structure} was proposed to infer dynamic diffusion probabilities based on information diffusion data and study the temporal evolution of information pathways. MMRate~\cite{wang2014mmrate} inferred multi-aspect transmission rates by incorporating aspect-level user interactions and various diffusion patterns. All above methods learned the probabilities from cascade sequences. Once a model is trained, it can be used for microscopic cascade prediction by simulating the generative process using Monte Carlo simulation.

Embedding-based approaches encoded each user into a parameterized real-valued vector and trained the parameters by maximizing an objective function. Embedded IC~\cite{bourigault2016representation} followed the pairwise independence assumption in IC model and modeled the diffusion probability between two users by a function of their user embeddings. Other embedding-based diffusion models~\cite{bourigault2014learning,gao2017novel} made even stronger assumptions that infected users are determined only by the source user and the content of information item. As shown in previous work~\cite{li2017deepcas}, such models with strong assumptions oversimplify the reality and generally show poor performance on real prediction tasks.

Existing deep-learning based microscopic cascade prediction approaches~\cite{hu2017will,wang2017topological} focused on the retweeting and sharing behaviors in a social network where ``who infected whom'' information is explicitly labeled in retweet chain. The next infected user candidates are also restricted to the neighboring users when the diffusion graph is known. However, the diffusion graph is usually unknown for most diffusion processes~\cite{wallinga2004different,myers2010convexity}. For example, during the contamination of a virus, by whom a patient gets infected is unobserved. Existing deep-learning based methods considered a much simpler problem and cannot be generalized to a general setting where the diffusion graph is unobserved. To the best of our knowledge, our work is the first attempt to employ deep learning techniques for general microscopic cascade prediction problem where the diffusion graph is unknown.
\subsection{Network Representation Learning}
Researchers have explored many algorithms to represent nodes in a network by real-valued vectors. By projecting topology structure into vectors, we can apply machine learning techniques for many network applications, \textit{e.g.} classification. Most network representation learning works focus on task unspecific learning where the downstream task is unknown. Early stage works~\cite{tang2009relational} use eigenvector computation to learn node embeddings. With the success of neural networks, people also employ simple neural networks for representation learning~\cite{perozzi2014deepwalk,tang2015line}. For task specific learning, a certain task such as classification~\cite{kipf2016semi} and recommendation~\cite{yang2017neural} is specified and the network embeddings serve as the bottom layer of their model as what we will do in this paper. In terms of diffusion prediction task, Embedded IC~\cite{bourigault2016representation} is proposed and will be used as our baseline method.
\section{Data Observation}
In this section, we will conduct data observation on real-world datasets and investigate the intrinsic relationships between activated users in a diffusion sequence. In specific, we will try to figure out whether consecutively activated users are more likely to be relevant and thus appear in more diffusion sequences together. We will first introduce the datasets.
\subsection{Datasets}
We collect four real-world cascade datasets that cover a variety of applications for evaluation. A cascade is an item or some kind of information that spreads through a set of users. Each cascade consists of a list of $(user,timestamp)$ pairs where each pair indicates the fact that the user gets infected at the timestamp.

\textbf{Lastfm} is a music streaming website. We collect the dataset from~\cite{celma2009music}. The dataset contains the full history of nearly $1,000$ users and the songs they listened to over one year. We treat each song as an item spreading through users and remove the users who listen to no more than $5$ songs.

\textbf{Irvine} is an online community for students at University of California, Irvine collected from~\cite{opsahl2009clustering}. Students can participate in and write posts on different forums. We regard each forum as an information item and remove the users who participate in no more than $5$ forums.

\textbf{Memetracker}~\footnote{http://www.memetracker.org} collects a million of news stories and blog posts and track the most frequent quotes and phrases, \textit{i.e.} memes, for studying the migration of memes across a group of people. Each meme is considered to be an information item and each URL of websites or blogs is regarded as a user. Following the settings of previous works~\cite{bourigault2016representation}, we filter the URLs to only keep the most active ones to alleviate the effect of noise.

\textbf{Twitter} dataset~\cite{hodas2014simple} concerns tweets containing URLs posted on Twitter during October 2010. The complete tweeting history of each URL is collected. We consider each distinct URL as a spreading item over Twitter users. We filter out the users with no more than $5$ tweets. Note that the scale of Twitter dataset is competitive and even larger than the datasets used in previous neural-based cascade modeling algorithms~\cite{bourigault2016representation,wang2017topological}.

Note that all the above datasets have no explicit evidence about by whom a user gets infected. Though we have the following relationship in Twitter dataset, we still cannot trace the source of by whom a user is encouraged to tweet a specific URL unless the user directly retweets.

\begin{figure*}[ht]
\centering
\begin{minipage}{\textwidth}
\subfigure[Lastfm]{
\includegraphics[width=0.23\textwidth]{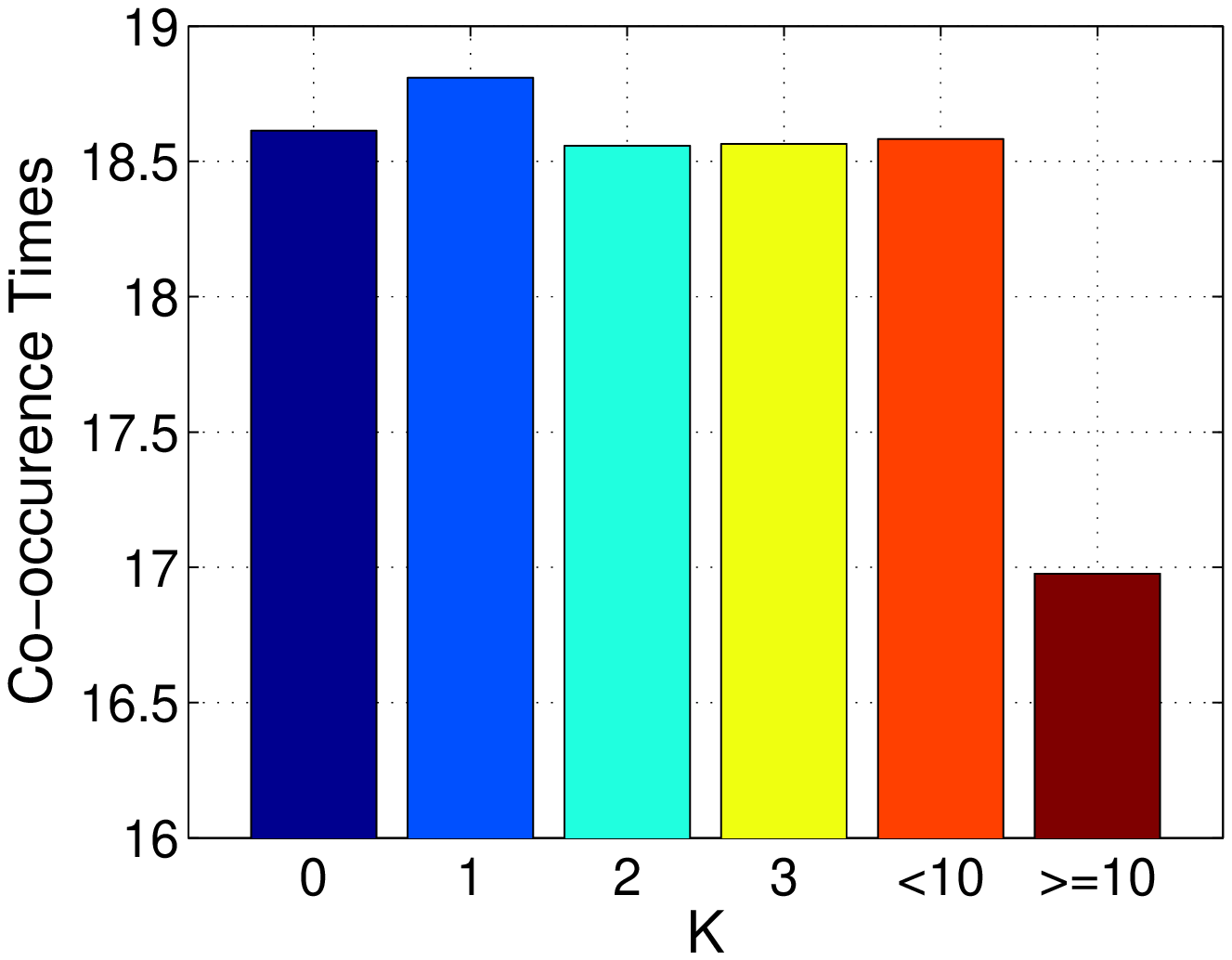}
}
\subfigure[Irvine]{
\includegraphics[width=0.23\textwidth]{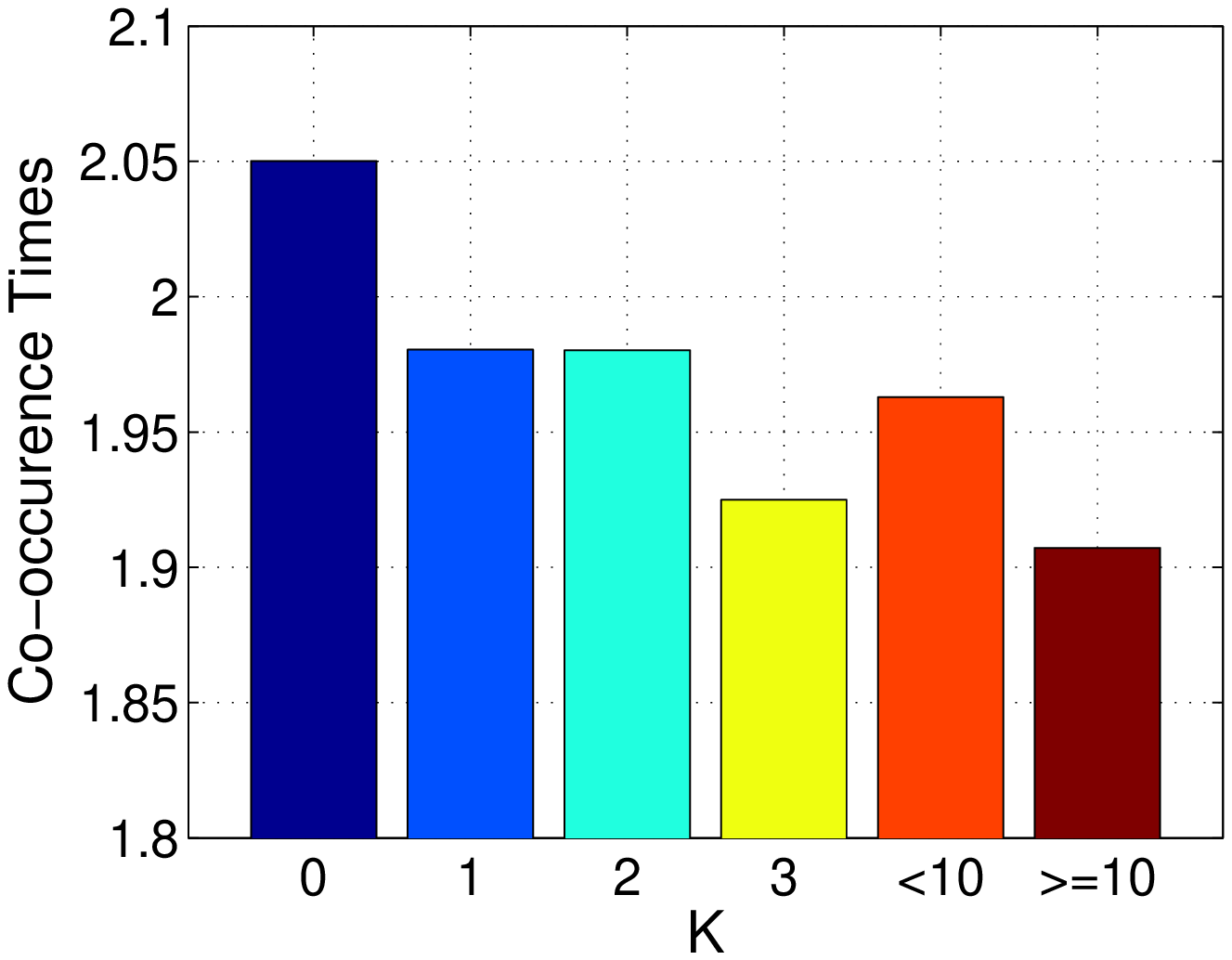}
}
\subfigure[Memetracker]{
\includegraphics[width=0.23\textwidth]{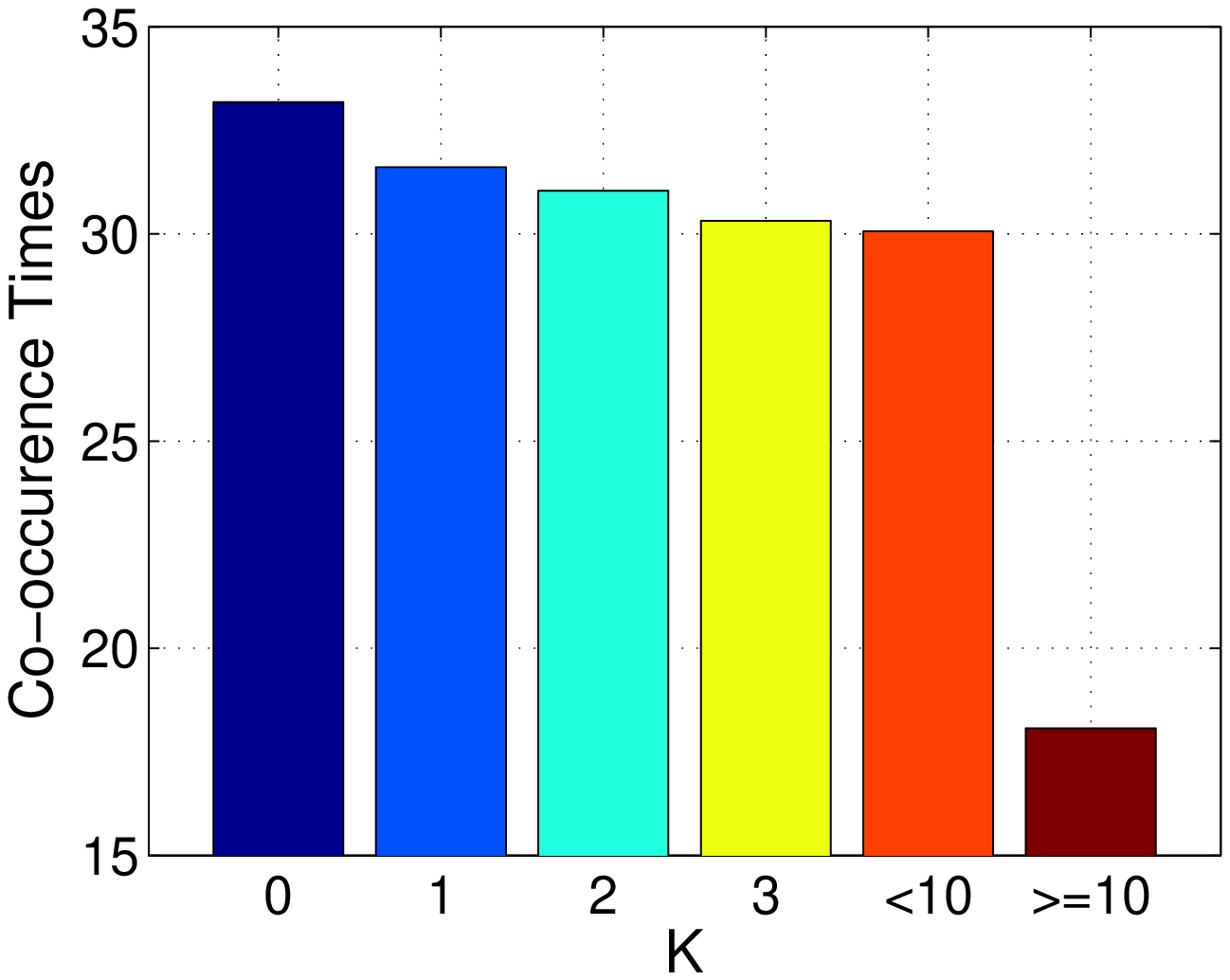}
}
\subfigure[Twitter]{
\includegraphics[width=0.23\textwidth]{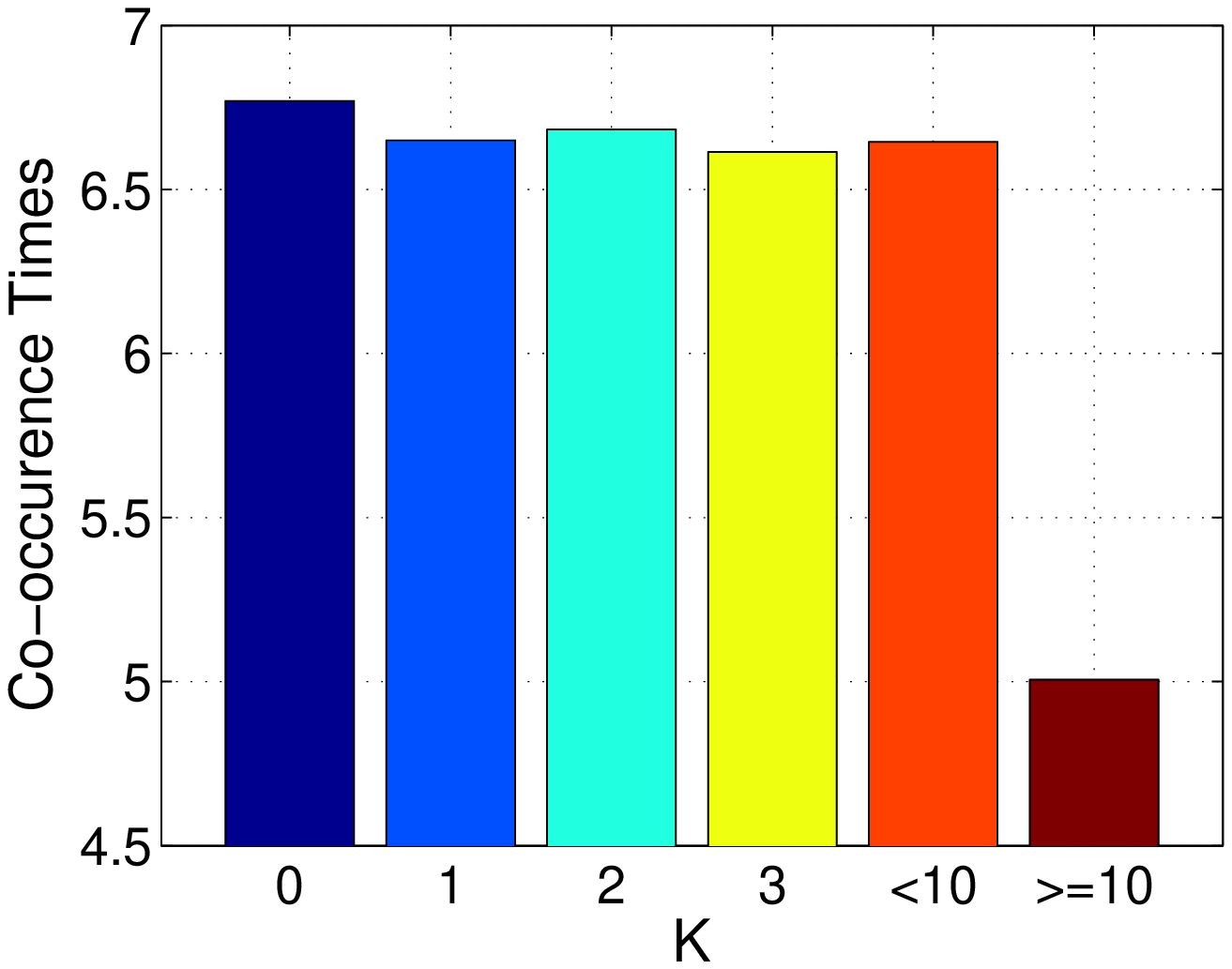}
}
\end{minipage}
\caption{Statistical results of the expectation of co-occurrence times of two random users given the fact that they are both infected by a cascade with K users infected between them. Here $<10$ and $\geq 10$ are average co-occurrence times for $K<10$ and $K\geq 10$.}
\label{fig:stat1}
\end{figure*}

\begin{figure*}[ht]
\centering
\begin{minipage}{\textwidth}
\subfigure[Lastfm]{
\includegraphics[width=0.23\textwidth]{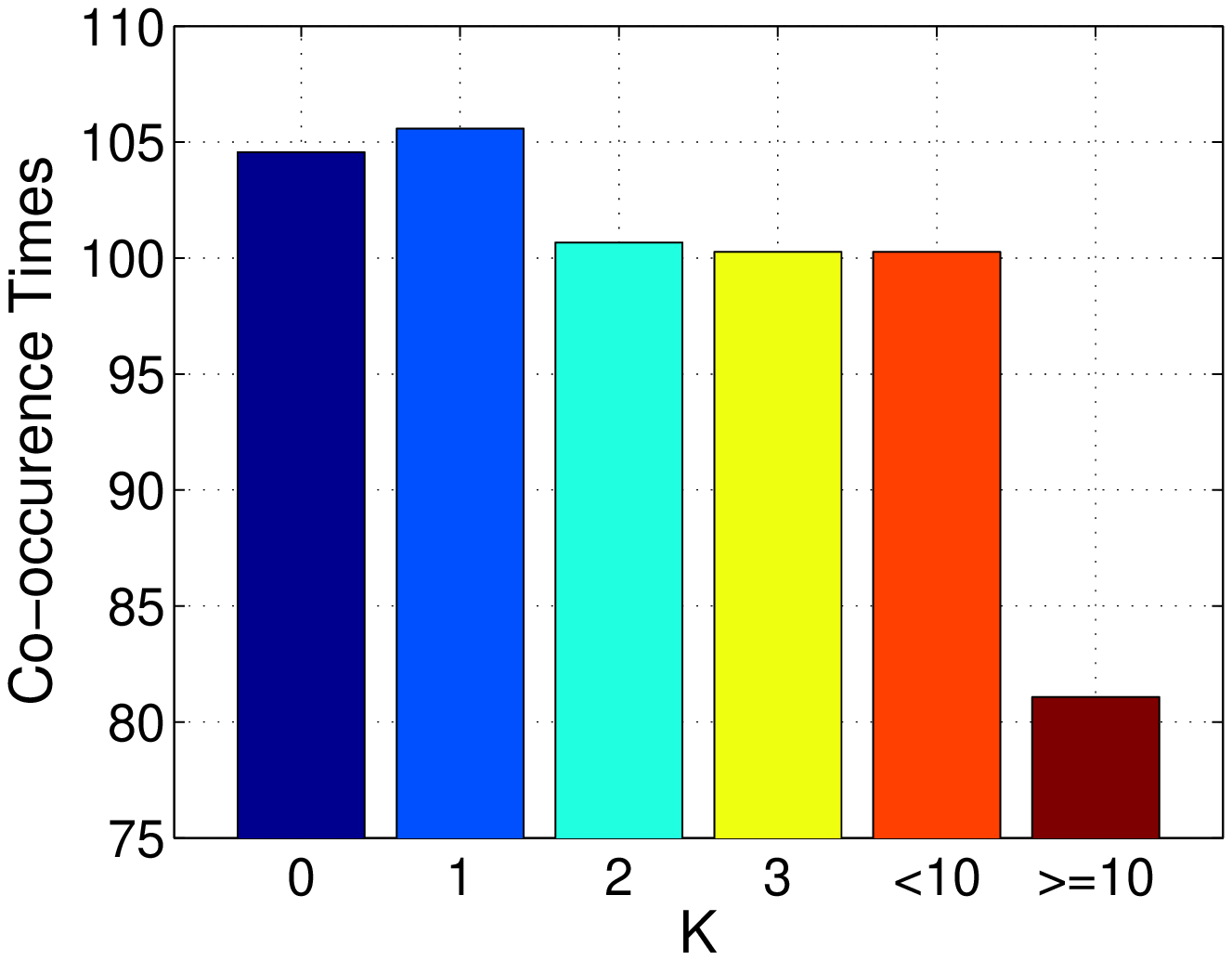}
}
\subfigure[Irvine]{
\includegraphics[width=0.23\textwidth]{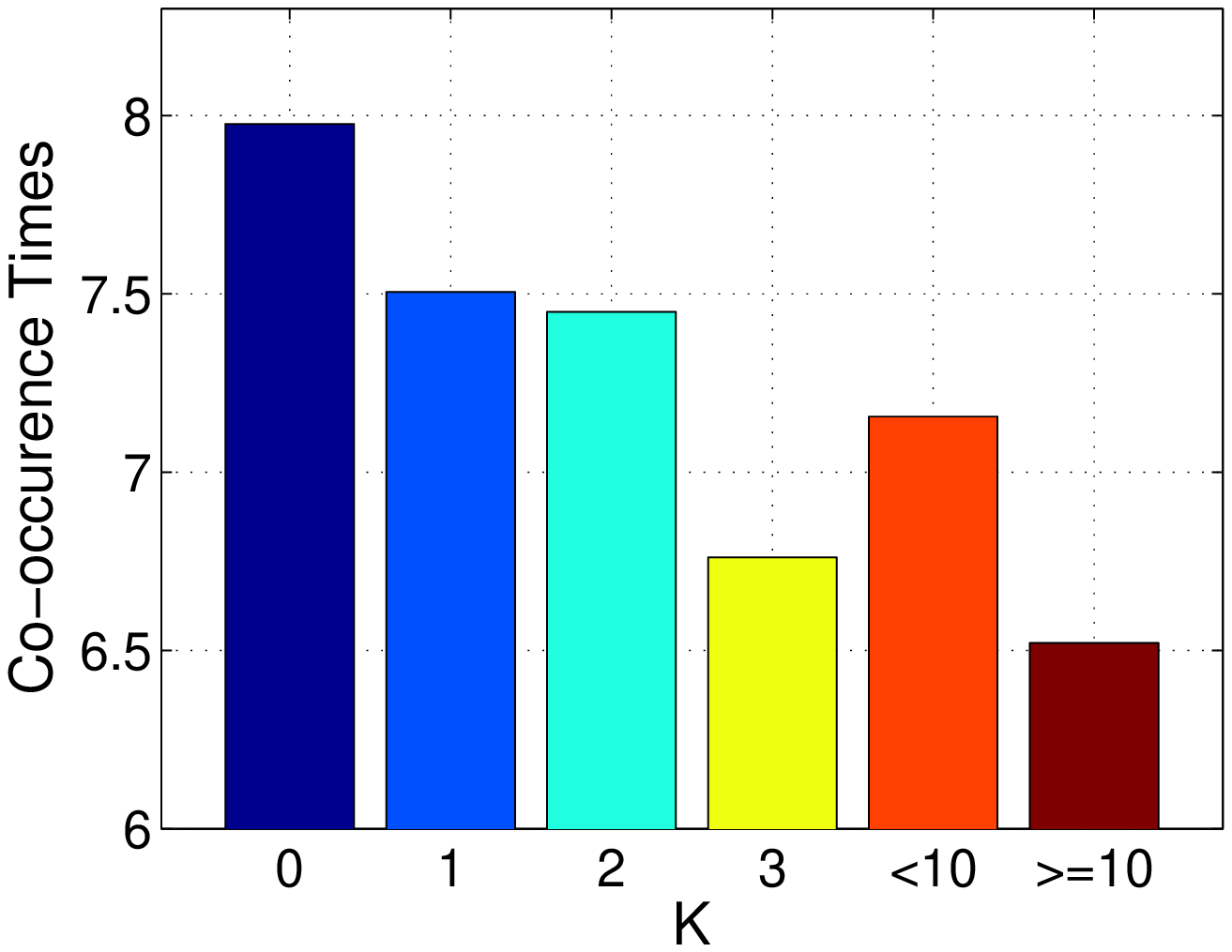}
}
\subfigure[Memetracker]{
\includegraphics[width=0.23\textwidth]{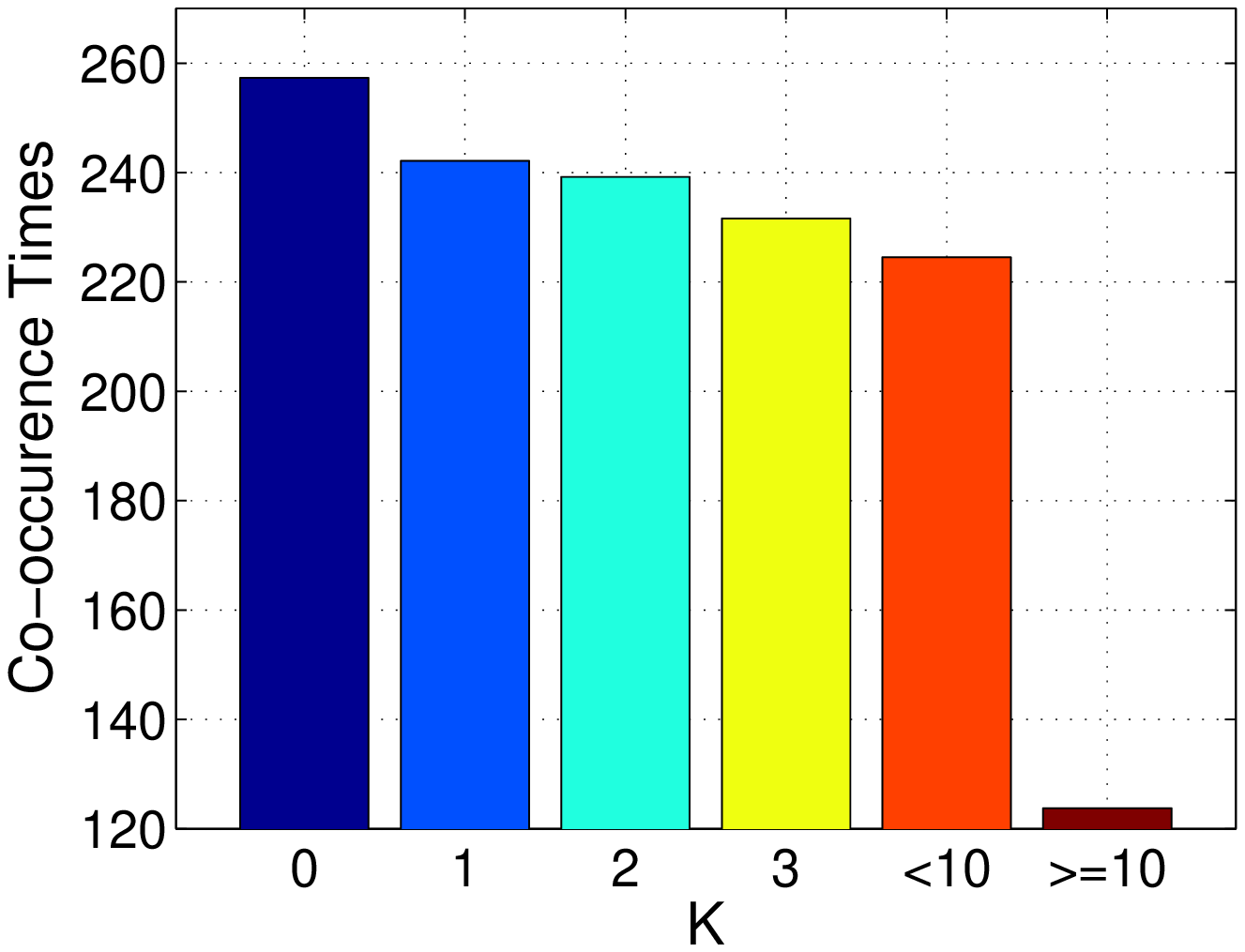}
}
\subfigure[Twitter]{
\includegraphics[width=0.23\textwidth]{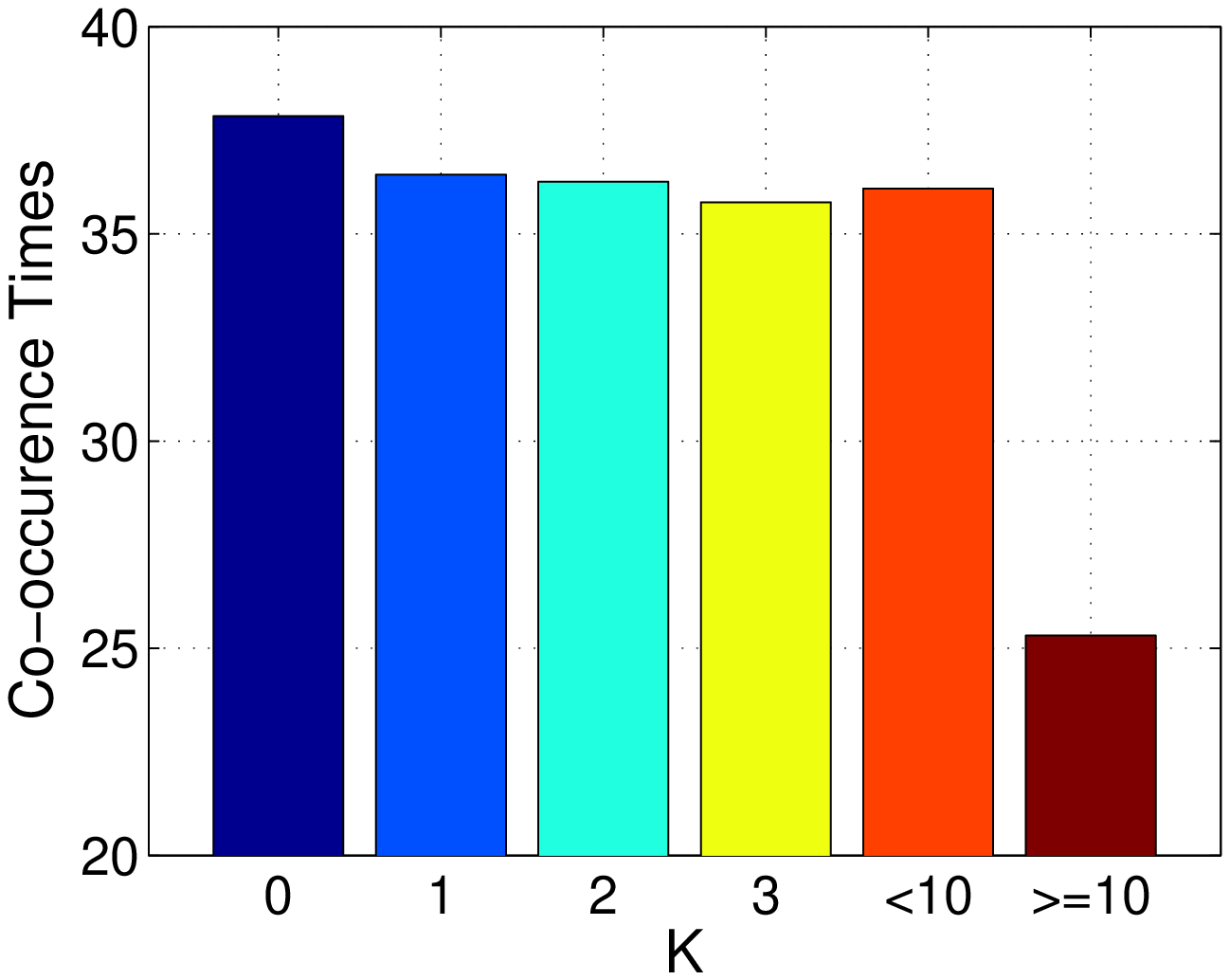}
}
\end{minipage}
\caption{Statistical results of the expectation of co-occurrence times of two random users given the fact that they are both infected by a cascade with K users infected between them and they are the \textbf{top 5\%} user pairs in terms of co-occurrence times satisfying the previous condition. }
\label{fig:stat2}
\end{figure*}

\begin{table}
  \caption{Statistics of Datasets.}
  \label{tab:statistics}
  \begin{tabular}{ccccc}
    \toprule
    Dataset & \# Users&\# Links& \#  Cascades& Avg. Length\\
    \midrule
    Lastfm & 982& 506,582& 23,802& 7.66\\
    Irvine & 540& 62,605& 471& 13.63\\
    Memetracker & 498& 158,194& 8,304& 8.43\\
    Twitter & 19,546 & 18,687,423& 6,158& 36.74\\
  \bottomrule
\end{tabular}
\end{table}

We list the statistics of datasets in Table~\ref{tab:statistics}. Since we have no interaction graph information between users, we assume that there exists a link between two users if they appear in the same cascade sequence. Each virtual ``link'' will be assigned a parameterized probability in traditional IC model and thus the space complexity of traditional methods is relatively high especially for large datasets. We also calculate the average cascade length of each dataset in the last column.
\subsection{Statistical Analysis}
Now we will try to reveal the correlation patterns between users by statistical results. By intuition, two consecutively infected users in a cascade sequence are more likely to have connections, \textit{e.g.} one infects another, and thus participate in many other diffusion sequences together.

To demonstrate this statement, we consider the following statistics: given the fact that user $u_i$ and $u_j$ are infected in a cascade sequence with $K$ users infected between them in this sequence, what will be the expectation of the number of cascade sequences that user $u_i$ and $u_j$ both participate in? Here $K=0$ indicates that user $u_i$ and $u_j$ are consecutively activated. If the intuition is true, then the expectation should decrease as $K$ increases.

Fig.~\ref{fig:stat1} presents the statistical results of all four datasets. Here we list the results for $K=0,1,2,3$ and the average for $K<10$ and $K\geq 10$. The statistics show that the expectations of co-occurrence times for $K<10$ are consistently larger than those for $K\geq 10$. Note that the gap is not very large for some datasets due to the long-tail effect. Therefore, we further present the results only for the top 5\% user pairs in terms of co-occurrence times for each $K$ in Fig.~\ref{fig:stat2}. We can see the differences more clearly in this setting.

These statistical results demonstrate that consecutively infected users in a cascade sequence are more likely to be relevant. By saying two users are ``relevant'', there could be a direct diffusion path between them or they are both likely to be infected by a third one. Also, we find that not only the most recently infected user will be relevant to the next infected one: As shown in Fig.~\ref{fig:stat1} and \ref{fig:stat2}, all recent infected users ($K=0,1,2,3$) could be relevant with minor differences (more relevant for smaller $K$). We will build our model based on these findings in next section.
\section{Method}
In this section, we will start by formalizing the problem and introducing the notations. Then we propose two heuristic assumptions according to the data observations as our basis and design a Neural Diffusion Model (NDM) using deep learning techniques. Finally, we will introduce the overall optimization function and other details of our model.
\subsection{Problem Formalization}
A cascade dataset records the information that an item spreads to whom and when during its diffusion. For example, the item could be a product and the cascade records who bought the product at what moment. However, in most cases, there exists no explicit interaction graph between the users~\cite{saito2008prediction,bourigault2016representation}. Therefore, we have no explicit information about how a user was infected by other users.

Formally, given user set $\mathcal{U}$ and observed cascade sequence set $\mathcal{C}$, each cascade $c_i \in \mathcal{C}$ consists a list of users $\{u_0^{i},u_1^{i}\dots u_{|c_i|-1}^{i}\}$ ranked by their infection time, where $|c_i|$ is the length of sequence $c_i$ and $u_j^{i}\in \mathcal{U}$ is the $j$-th user in the sequence $c_i$. Note that we only consider the order of users getting infected and ignore the exact timestamps of infections in this paper as previous works did~\cite{bourigault2016representation,kefato2017di,wang2017topological}.

In this paper, our goal is to learn a cascade prediction model which can predict the next infected user $u_{j+1}$ given a partially observed cascade sequence $\{u_0,u_1\dots u_j\}$. The learned model is able to predict the entire infected user sequence based on the first few observed infected users and thus be used for microscopic cascade prediction illustrated in Figure~\ref{fig:intro}. In our model, we add a virtual user called ``Terminate'' to the user set $\mathcal{U}$. At training phase, we append ``Terminate'' to the end of each cascade sequence and allow the model to predict next infected user as ``Terminate'' to indicate that no more users will be infected in this cascade.

Further, we represent each user by a parameterized real-valued vector to project users into vector space. The real-valued vectors are also called embeddings. We denote the embedding of user $u$ as $emb(u)\in \mathbb{R}^d$ where $d$ is the dimension of embeddings. In our model, a larger inner product between the embeddings of two users indicates a stronger correlation between the users.

\subsection{Model Assumptions}
In traditional Independent Cascade (IC) model~\cite{kempe2003maximizing} settings, all previously infected users can activate a new user independently and equally regardless of their orders of getting infected. Many extensions of IC model further considered time delay information such as continuous time IC (CTIC)~\cite{saito2009learning} and Netrate~\cite{rodriguez2014uncovering}. However, none of these models tried to find out which users are actually active and more likely to activate other users at the moment. To address this issue, we propose the following assumption.

\textbf{Assumption 1.} Given a recently infected user $u$, users that are strongly correlated to user $u$ including user $u$ itself are more likely to be active.

This assumption is intuitive and straight-forward. As a newly activated user, $u$ should be active and may infect other users. The users strongly correlated to user $u$ are probably the reason why user $u$ gets activated recently and thus more likely to be active than other users at the moment. We further propose the concept of ``active user embedding'' to characterize all such active users.

\textbf{Definition 1.} For each recently infected user $u$, we aim to learn an active user embedding $act(u)\in \mathbb{R}^d$ which represents the embedding of all active users related to user $u$, and can be used for predicting the next infected user in next step.

The active user embedding $act(u_j)$ characterizes the potential active users related to the fact that user $u_j$ gets infected. From the data observations, we can see that all recently infected users could be relevant to the next infected one. Therefore, the active user embeddings of all recently infected users should contribute to the prediction of next infected user, which leads to the following assumption.

\textbf{Assumption 2.} All recently infected users should contribute to the prediction of next infected user and be processed differently according to the order of getting infected.

Compared with the strong assumptions made by IC-based and embedding-based method introduced in related works, our heuristic assumptions allow our model to be more flexible and better fit cascade data. Now we will introduce how to build our model based on these two assumptions, \textit{i.e.} extracting active users and unifying these embeddings for prediction.
\subsection{Extracting Active Users with Attention Mechanism}
\begin{figure*}[htb]
\centering
\includegraphics[width=.85\linewidth]{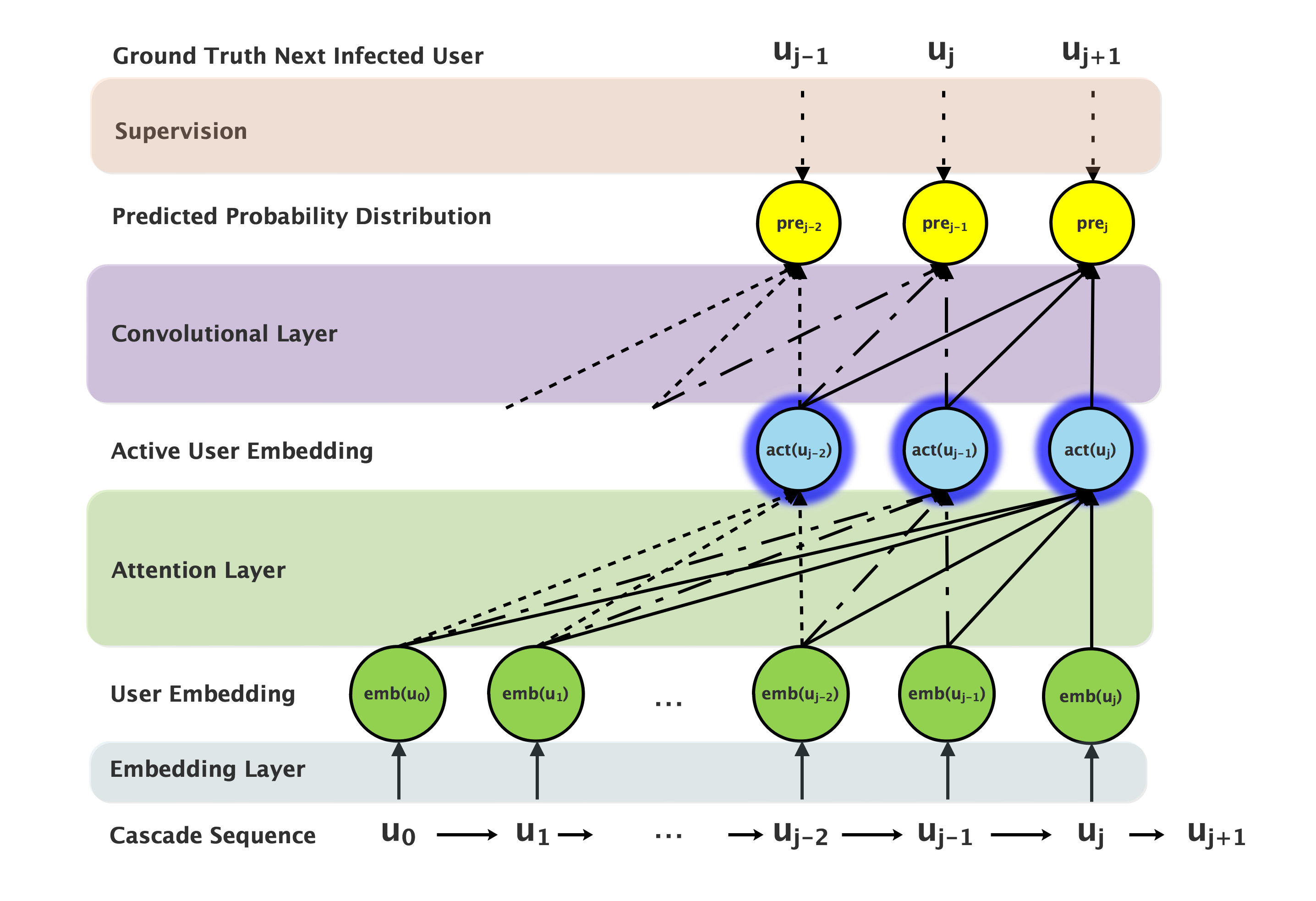}
\caption{An overview of our Neural Diffusion Model (NDM). NDM sequentially predicts the next infected user based on the active embeddings (the blue nodes) of recent activated users and the active embeddings is computed by an attention layer over the user embeddings (the green nodes) of all previous infected users.}
\label{fig:model}
\end{figure*}

For the purpose of computing active user embeddings, we propose to use attention mechanism~\cite{bahdanau2014neural,vaswani2017attention} to extract the most likely active users by giving them more weights than other users. As shown in Figure~\ref{fig:model}, the active embedding of user $u_j$ is computed as a weighted sum of previously infected users:
\begin{equation}
act(u_j)=\sum_{k=0}^j w_{jk}emb(u_k),
\label{eq:active}
\end{equation}
where the weight of $u_k$ is
\begin{equation}
w_{jk}=\frac{\exp(emb(u_j) emb(u_k)^T)}{\sum_{m=0}^j \exp(emb(u_j) emb(u_m)^T)}.
\end{equation}
Note that $w_{jk}\in (0,1)$ for every $k$ and $\sum_{m=0}^j w_{jm}=1$. $w_{jk}$ is the normalized inner product between the embeddings of $u_j$ and $u_k$ which indicates the strength of correlation between them.

From the definition of active user embedding $act(u_j)$ in Eq.~\ref{eq:active}, we can see that the user embeddings $emb(u_k)$ which have a larger inner product with $emb(u_j)$ will be allocated a larger weight $w_{jk}$. This formula naturally follows our assumption that users strongly correlated to user $u$ including user $u$ itself should be paid more attention.

To fully utilize the advantages of a neural model, we further employ the multi-head attention~\cite{vaswani2017attention} to improve the expressibility. Multi-head attention projects the user embeddings into multiple subspaces with different linear projections. Then multi-head attention performs attention mechanism on each subspace independently. Finally, multi-head attention concatenates the attention embeddings in all subspaces and feeds the result into a linear projection again.

Formally, in a multi-head attention with $h$ heads, the embedding of $i$-th head is computed as
\begin{equation}
head_i=\sum_{k=0}^j w_{jk}^i emb(u_k)W_i^V,
\end{equation}
where
\begin{equation}
w_{jk}^i=\frac{\exp(emb(u_j)W_i^Q (emb(u_k)W_i^K)^T)}{\sum_{m=0}^j \exp(emb(u_j)W_i^Q (emb(u_m)W_i^K)^T)},
\label{multihead}
\end{equation}
$W_i^V,W_i^Q,W_i^K \in \mathbb{R}^{d\times d}$ are head-specific linear projection matrices. In particular, $W_i^Q$ and $W_i^K$ can be seen to project user embeddings into \textit{receiver} space and \textit{sender} space respectively for asymmetric modeling.

Then we have the active user embedding $act(u_j)$
\begin{equation}
act(u_j)=[head_1,head_2\dots head_h]W^O,
\end{equation}
where $[]$ indicates concatenation operation and $W^O\in \mathbb{R}^{hd\times d}$ projects the concatenated results into $d$-dimensional vector space.

Multi-head attention allows the model to ``divide and conquer'' information from different perspectives (\textit{i.e.} subspaces) independently and thus is more powerful than the traditional attention mechanism.
\subsection{Unifying Active User Embeddings for Prediction with Convolutional Network}
Different from previous works~\cite{rodriguez2011uncovering,rodriguez2014uncovering} which directly give a time-decay weight that assumes larger weights for the most recently infected users, we propose to use a parameterized neural network to handle the active user embeddings at different positions. Compared with a predefined exponential-decay weighting function~\cite{rodriguez2014uncovering}, a parameterized neural network can be learned automatically to fit the real-world dataset and capture the intrinsic relationship between active user embedding at each position and next infected user prediction. In this paper, we consider Convolutional Neural Network (CNN) to meet this purpose.

CNN has been widely used in image recognition~\cite{lecun2015lenet}, recommender systems~\cite{van2013deep} and natural language processing~\cite{collobert2008unified}. CNN is a \textit{shift-invariant} neural network and allows us to assign position-specific linear projections to the embeddings.

Figure~\ref{fig:model} illustrates an example where the window size of our convolutional layer $\textit{win}=3$. The convolutional layer first converts each active user embedding $act(u_{j-n})$ into a $|\mathcal{U}|$-dimensional vector by a position-specific linear projection matrix $W_n^C\in \mathbb{R}^{d\times |\mathcal{U}|}$ for $n=0,1\dots \textit{win}-1$. Then the convolutional layer sums up the projected vectors and normalizes the summation by \textit{softmax} function.

Formally, given partially observed cascade sequence $(u_0,u_1\dots u_j)$, the predicted probability distribution $pre_j\in \mathbb{R}^{|\mathcal{U}|}$ is
\begin{equation}
pre_j=\text{softmax}(\sum_{n=0}^{\textit{win}-1} act(u_{j-n})W_n^C),
\label{eq:softmax}
\end{equation}
where $\text{softmax}(x)[i]=\frac{\exp(x[i])}{\sum_p \exp(x[p])}$ and $x[i]$ denotes the $i$-th entry of a vector $x$. Each entry of $pre_j$ represents the probability that the corresponding user gets infected at next step. 

Since the initial user $u_0$ plays an important role in the whole diffusion process, we further take $u_0$ into consideration:
\begin{equation}
pre_j=\text{softmax}(\sum_{n=0}^{\textit{win}-1} act(u_{j-n})W_n^C+act(u_0)W_{init}^C\cdot F_{init}),
\label{eq:softmax_init}
\end{equation}
where $W_{init}^C\in \mathbb{R}^{d\times |\mathcal{U}|}$ is the projection matrix for initial user $u_0$ and $F_{init}\in \{0,1\}$ is a hyperparameter which controls whether incorporate initial user for prediction or not.

\subsection{Overall Architecture, Model Details and Learning Algorithms}
We naturally maximize the log-likelihood of all observed cascade sequences to build the overall optimization function.
\begin{equation}
\mathcal{L}(\Theta)=\sum_{c_i\in \mathcal{C}}\sum_{j=0}^{|c_i|-2} \log pre_j^i[u_{j+1}^i],
\end{equation}
where $pre_j^i[u_{j+1}^i]$ is the predicted probability of ground truth next infected user $u_{j+1}^i$ at position $j$ in cascade $c_i$, and $\Theta$ is the set of all parameters need to be learned, including user embeddings $emb(u)\in \mathbb{R}^d$ for each $u\in \mathcal{U}$, projection matrices in multi-head attention $W_n^V,W_n^Q,W_n^K\in \mathbb{R}^{d\times d}$ for $n=1,2\dots h, W^O\in \mathbb{R}^{hd\times d}$ and projection matrices in convolutional layer $W_{init}^C,W_n^C\in \mathbb{R}^{d\times |\mathcal{U}|}$ for $n=0,1\dots \text{win}-1$.

\textbf{Implementation Details.} We implement our model using PyTorch~\footnote{http://pytorch.org} and optimize the parameters by gradient descent with Adam optimizer~\cite{kingma2014adam}. We further employ layer normalization~\cite{ba2016layer} and residual connection~\cite{he2016deep} operation to active user embedding to avoid gradient explosion or vanishment problem that may occur in deep neural networks. In other words, the active user embedding $act(u)$ is replaced by $LayerNorm(emb(u)+act(u))$ instead where the $LayerNorm(\cdot)$ function encourages the output to have zero mean and unit variance. We also use dropout~\cite{srivastava2014dropout} to the attention mechanism to prevent our model from overfitting and the dropout rate is set to $0.1$. Since the same user will not be infected twice, we mask the users that are already infected in the Eq.~\ref{eq:softmax_init} so that they won't be predicted. We release our source code at \url{github}~\footnote{https://github.com/albertyang33/NeuralDiffusionModel} and all the details are listed. Hyperparameter settings will be introduced in next section.

\textbf{Complexity.} The space complexity of our model is $O(d|\mathcal{U}|)$ where $d$ is the embedding dimension which is much less than the size of user set. Note that the space complexity of training traditional IC model will go up to $O(|\mathcal{U}|^2)$ because we need to assign an infection probability between each pair of potential linked users. Therefore, the space complexity of our neural model is less than that of traditional IC methods.

The computation of a single active embedding takes $O(|c_i|d^2)$ time where $c_i$ is the length of corresponding cascade and the next infected user prediction in Eq.~\ref{eq:softmax_init} step takes $O(d|\mathcal{U}|)$ time. Hence the time complexity of training a single cascade is $O(\sum_{c_i\in \mathcal{C}}(|c_i|^2d^2+|c_i|d|\mathcal{U}|))$ which is competitive with previous neural-based models such as embedded IC model~\cite{bourigault2016representation}. But as we will show in the experiments, our model converges much faster than embedded IC model and is capable of handling large-scale dataset.

\section{Experiments}
We conduct experiments on diffusion prediction task as previous works did~\cite{bourigault2016representation} to evaluate the performance of our model and various baseline methods. We will first introduce the baseline methods, evaluation metrics and hyperparameter settings. Then we will present the experimental results and give further analysis about the evaluation.

\subsection{Baselines}
We consider a number of state-of-the-art baselines to demonstrate the effectiveness of our algorithm. Most of the baseline methods will learn a transition probability matrix $M\in \mathbb{R}^{|\mathcal{U}|\times|\mathcal{U}|}$ from cascade sequences where each entry $M_{ij}$ represents the probability that user $u_j$ gets infected by $u_i$ when $u_i$ is activated.

\textbf{Netrate}~\cite{rodriguez2014uncovering} considers the time-varying dynamics of diffusion probability through each link and defines three transmission probability models, \textit{i.e.} exponential, power-law and Rayleigh, which encourage the diffusion probability to decrease as the time interval increases. In our experiments, we only report the results of exponential model since the other two models give similar results.

\textbf{Infopath}~\cite{gomez2013structure} also targets on inferring dynamic diffusion probabilities based on information diffusion data. Infopath employs stochastic gradient to estimate the temporal dynamics and studies the temporal evolution of information pathways.

\textbf{Embedded IC}~\cite{bourigault2016representation} explores representation learning technique and models the diffusion probability between two users by a function of their user embeddings instead of a static value. Embedded IC model is trained by stochastic gradient descent method.

\textbf{LSTM} is a widely used neural network framework~\cite{hochreiter1997long} for sequential data modeling and has been used for cascade modeling recently. Previous works employ LSTM for some simpler tasks such as popularity prediction~\cite{li2017deepcas} and cascade prediction with known diffusion graph~\cite{wang2017topological,hu2017will}. Since none of these works are directly comparable to ours, we adopt LSTM network for comparison by adding a softmax classifier to the hidden state of LSTM at each step for next infected user prediction.
\subsection{Hyperparameter Settings for Neural Models}
Though the parameter space of neural network based methods is much less than that of traditional IC models, we have to set several hyperparameters to train neural models. To tune the hyperparameters, we randomly select $10\%$ of training cascade sequences as validation set. Note that all training cascade sequences including the validation set will be used to train the final model for testing.

\begin{table*}
\centering
  \caption{Experimental results on diffusion prediction.}
  \label{tab:main}
  \begin{tabular}{|c|c|p{1.5cm}<{\centering}|p{1.5cm}<{\centering}|p{1.8cm}<{\centering}|p{1.5cm}<{\centering}|p{1.5cm}<{\centering}|c|}
 \hline
 \multirow{2}{*}{Metric} &\multirow{2}{*}{Dataset}&
 \multicolumn{5}{c|}{Method} &\multirow{2}{*}{Improvement}\\
 \cline{3-7} &
   & Netrate & Infopath & Embedded IC & LSTM & NDM &\\
 \hline
 \multirow{4}{*}{Macro-F1} & Lastfm &
  0.017 & 0.030 & 0.020 & 0.026& \textbf{0.056} & +87\%\\
 \cline{2-8} & Memetracker & 0.068 & 0.110 & 0.060 & 0.102 & \textbf{0.139} & +26\%\\
 \cline{2-8} & Irvine & 0.032 & 0.052 & 0.054 & 0.041 & \textbf{0.076} & +41\%\\
 \cline{2-8} & Twitter & - & 0.044 & - & 0.103 & \textbf{0.139} & +35\%\\
 \hline
 \multirow{4}{*}{Micro-F1} & Lastfm &
 0.007 & 0.046 & 0.085 & 0.072 & \textbf{0.095}& +12\% \\
 \cline{2-8} & Memetracker & 0.050 & 0.142 & 0.115 & 0.137 & \textbf{0.171} & +20\%\\
 \cline{2-8} & Irvine & 0.029 & 0.073 & 0.102 & 0.080 & \textbf{0.108} & +6\%\\
 \cline{2-8} & Twitter & - & 0.010 & - & 0.052 & \textbf{0.087} & +67\%\\
 \hline
 \end{tabular}
\end{table*}

\begin{table*}
\centering
  \caption{Experimental results on diffusion prediction at early stage where only the first five infected users are predicted in each cascade.}
  \label{tab:main5}
  \begin{tabular}{|c|c|p{1.5cm}<{\centering}|p{1.5cm}<{\centering}|p{1.8cm}<{\centering}|p{1.5cm}<{\centering}|p{1.5cm}<{\centering}|c|}
 \hline
 \multirow{2}{*}{Metric} &\multirow{2}{*}{Dataset}&
 \multicolumn{5}{c|}{Method} &\multirow{2}{*}{Improvement}\\
 \cline{3-7} &
   & Netrate & Infopath & Embedded IC & LSTM & NDM &\\
 \hline
 \multirow{4}{*}{Macro-F1} & Lastfm &
  0.018 & 0.028 & 0.010 & 0.018& \textbf{0.048} & +71\%\\
 \cline{2-8} & Memetracker & 0.071 & 0.094 & 0.042 & 0.091 & \textbf{0.122} & +30\%\\
 \cline{2-8} & Irvine & 0.031 & 0.030 & 0.027 & 0.018 & \textbf{0.064}& +106\%\\
 \cline{2-8} & Twitter & - & 0.040 & - & 0.097 & \textbf{0.123} & +27\% \\
 \hline
 \multirow{4}{*}{Micro-F1} & Lastfm &
 0.016 & 0.035 & 0.013 & 0.019 & \textbf{0.045}& +29\%\\
 \cline{2-8} & Memetracker & 0.076 & 0.106 & 0.040 & 0.094 & \textbf{0.126} & +19\%\\
 \cline{2-8} & Irvine & 0.028 & 0.030 & 0.029 & 0.020 &\textbf{0.065} & +117\%\\
 \cline{2-8} & Twitter & - & 0.050 & - & 0.093 & \textbf{0.118} & +27\%\\
 \hline
 \end{tabular}
\end{table*}

For Embedded IC model, the dimension of user embeddings is selected from $\{25,50,100\}$ as the original paper did~\cite{bourigault2016representation}. For LSTM model, the dimensions of user embeddings and hidden states are set to the best choice from $\{16,32,64,128\}$. For our model NDM, the number of heads used in multi-head attention is set to $h=8$, the window size of convolutional network is set to $\textit{win}=3$ and the dimension of user embeddings is set to $d=64$. Note that we use the same set of $(h,\textit{win},d)$ for all the datasets. The flag $F_{init}$ in Eq.~\ref{eq:softmax_init} which determines whether the initial user is used for prediction is set to $F_{init}=1$ for Twitter dataset and $F_{init}=0$ for the other three datasets. We will show the robustness of our model in parameter sensitivity subsection.

Note that neural models, \textit{i.e.} Embedded IC, LSTM and NDM, are based on matrix multiplication operations and thus naturally benefit from the GPU acceleration. Therefore, we train these three methods on a GPU device (GeForce GTX TITAN X) instead of a CPU device (Intel Xeon E5-2620 @ 2.0GHz).
\subsection{Diffusion Prediction}
To compare the ability of cascade modeling, we evaluate our model and all baseline methods on the diffusion prediction task. We follow the experimental settings in Embedded IC~\cite{bourigault2016representation}. We randomly select $90\%$ cascade sequences as training set and the rest as test set. For each cascade sequence $c=(u_0,u_1,u_2\dots)$ in the test set, only the initial user $u_0$ is given and all successively infected users $G^c=\{u_1,u_2\dots u_{|G^c|}\}$ need to be predicted.

All baseline methods and our model are required to predict a set of users and the results will be compared with ground truth infected user set $G$. For baseline methods that ground in IC model, \textit{i.e.} Netrate, Infopath and Embedded IC, we will simulate the infection process according to the learned pairwise diffusion probability and their corresponding generation process. For LSTM and our model, we can sequentially sample a user according to the probability distribution of softmax classifier at each step.

Note that the ground truth infected user set could also be partially observed because the datasets are crawled within a short time window. Therefore, for each test sequence $c$ with $|G^c|$ ground truth infected users, all the algorithms are only required to predict the first $|G^c|$ infected users in a single simulation. Also note that the simulation may terminate and stop infecting new users before activating $|G^c|$ users.

We conduct $1000$ times Monte Carlo simulations for each test cascade sequence $c$ for all algorithms and compute the infection probability $P_u^c$ of each user $u\in \mathcal{U}$. We evaluate the prediction results using two classic evaluation metrics: Macro-F1 and Micro-F1.

\textbf{Macro-F1.} Macro-averaged F1 first computes the precision $pre_c$, recall $rec_c$ and F1 score $f_c$ locally for each test cascade sequence $c$ in the test set $\mathcal{C}_{test}$. Then macro-averaged F-measure takes the average over all test cascade sequences:
\begin{displaymath}
pre_c=\frac{\sum_{u\in G^c}P_u^c}{\sum_{u\in \mathcal{U}} P_u^c},rec_c=\frac{\sum_{u\in G^c}P_u^c}{|G^c|},f_c=\frac{2pre_c\cdot rec_c}{pre_c+rec_c},
\end{displaymath}
\begin{displaymath}
Macro-F1=\frac{\sum_{c\in |\mathcal{C}_{test}|}f_c}{|\mathcal{C}_{test}|}.
\end{displaymath}

\textbf{Micro-F1.} Micro-averaged F1 computes precision $pre$, recall $rec$ globally by averaging over all predictions and serves as a complementary view by giving larger weights to longer cascades:
\begin{displaymath}
pre=\frac{\sum_{c\in |\mathcal{C}_{test}|}\sum_{u\in G^c}P_u^c}{\sum_{c\in |\mathcal{C}_{test}|}\sum_{u\in \mathcal{U}} P_u^c},rec=\frac{\sum_{c\in |\mathcal{C}_{test}|}\sum_{u\in G^c}P_u^c}{\sum_{c\in |\mathcal{C}_{test}|}|G^c|},
\end{displaymath}
\begin{displaymath}
Micro-F1=\frac{2pre\cdot rec}{pre+rec}.
\end{displaymath}

To further evaluate the performance of cascade prediction at early stage, we conduct additional experiments by only predicting the first five infected users in each test cascade. We present the experimental results in Table~\ref{tab:main} and~\ref{tab:main5}. Here ``-'' indicates that the algorithm fails to converge in $72$ hours. The last column represents the relative improvement of NDM against the best performing baseline method. We have the following observations:

(1) NDM consistently and significantly outperforms all the baseline methods. As shown in Table~\ref{tab:main}, the relative improvement against the best performing baseline is at least $26\%$ in terms of Macro-F1 score. The improvement on Micro-F1 score further demonstrates the effectiveness and robustness of our proposed model. The results also indicate that well-designed neural network models are able to surpass traditional cascade methods on cascade modeling.

(2) NDM has even more significant improvements on cascade prediction task \textit{at early stage}. As shown in Table~\ref{tab:main5}, NDM outperforms all baselines by a large margin on both Macro and Micro F1 scores. Note that it's very important to predict the first wave of infected users accurately for real-world applications because a wrong prediction will lead to error propagation in following stages. A precise prediction of infected users at early stage enables us to better control the spread of information items through users. For example, we can prevent the spread of a rumor by warning the most vulnerable users in advance and promote the spread of a product by paying the most potential customers more attention. This experiment demonstrates that NDM has the ability to be used for real-world applications.

(3) NDM is capable of handling large-scale cascade datasets. Previous neural-based method, Embedded IC, fails to converge in $72$ hours on Twitter dataset with around $20$ thousand users and $19$ million of potential links. In contrast, NDM converges in $6$ hours on this dataset with the same GPU device, which is at least $10$ times faster than Embedded IC. This observation demonstrates the scalability of NDM.
\subsection{Social Link Prediction}
Sometimes the underlying social network of users is available, \textit{e.g.} the Twitter dataset used in our experiments. In the Twitter dataset, a network of Twitter followers is observed though the information diffusion is not necessarily passed through the edges of the social network. Though the diffusion network and the social network do not strictly align with each other, we still expect that the most closely related users in information diffusion should also be  socially connected in the social network. Therefore, we conduct social link prediction experiments to verify this statement.

Firstly, we need to specify ``the most closely related users''. For Infopath algorithm, the model output directly contains the diffusion probability of each inferred edge. Thus we can rank the users by their diffusion probability to get the most closely related users. For LSTM and our model NDM, the output contains users' embeddings as real-valued vectors and we can simply use the inner product or the multi-head attention weight in Eq.~\ref{multihead} between user embeddings to measure the closeness of two users.

\begin{figure}[tb]
\centering
\includegraphics[width=.99\columnwidth]{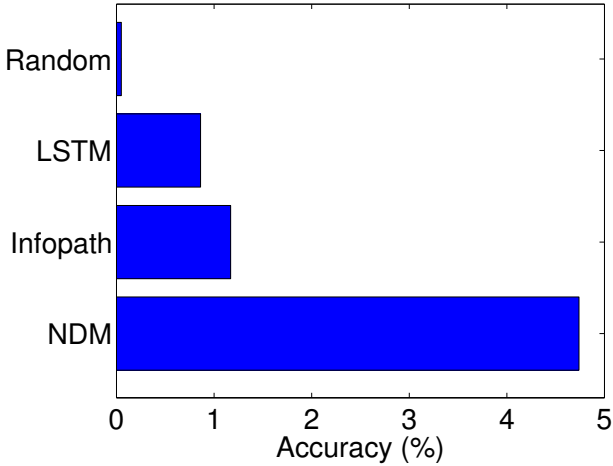}
\caption{Experimental results on social link prediction in Twitter dataset.}
\label{fig:net1}
\end{figure}

Secondly, we use the following experimental settings for evaluation. Note that this setting is a reasonable choice but not the only choice. For each user $u$ in the dataset, we select the most closely related user according to the first step and check whether this most closely related user is a follower of user $u$ or not. Then we naturally use the accuracy as evaluation metric. We present the experimental results in Fig.~\ref{fig:net1}.

\begin{figure}[tb]
\centering
\includegraphics[width=.99\columnwidth]{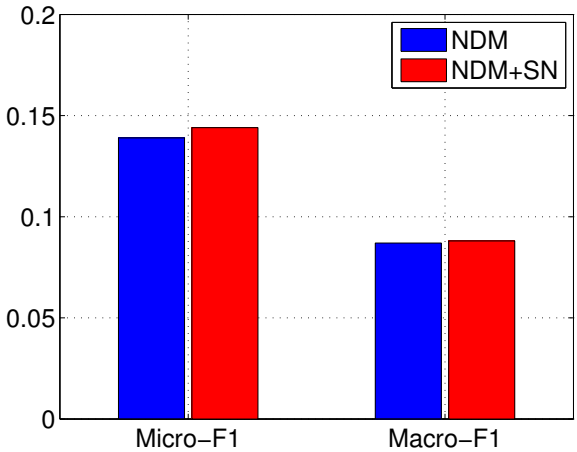}
\caption{Comparisons between NDM and NDM+SN on diffusion prediction.}
\label{fig:net2}
\end{figure}
\begin{figure}[tb]
\centering
\includegraphics[width=.99\columnwidth]{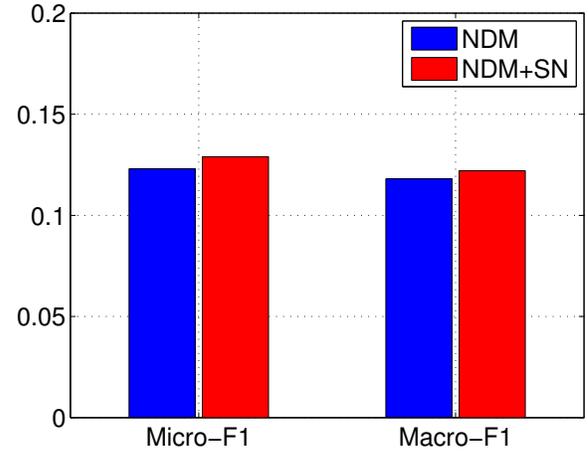}
\caption{Comparisons between NDM and NDM+SN on diffusion prediction at early stage where only the first five infected users are predicted.}
\label{fig:net3}
\end{figure}

From the results in Fig.~\ref{fig:net1} we can see that all three algorithms, \textit{i.e.} LSTM, Infopath and NDM, are able to predict the social links in Twitter dataset to some extent. Even the accuracy of LSTM is much higher than a random guess (around $0.005\%$). This indicates that information spreads through some social links frequently and thus these links can be inferred successfully. Moreover, our NDM model performs best on this task. This fact indicates that NDM can better capture the intrinsic relationship between users. Also, the absolute value of social link prediction accuracy is still not high enough (less than $5\%$). One possible reason is that the overlap between a diffusion network and a social network is small compared with the entire network.
\subsection{Benefits from Social Network}
On the other side, we also hope that diffusion prediction process could benefit from the observed social network structure. We apply a simple modification on our NDM model to take advantage of the social network. Now we will introduce the modification in detail.

\begin{figure*}[ht]
\centering
\begin{minipage}{\textwidth}
\subfigure[\# heads $h\in\{1,2,4,8\}$]{
\includegraphics[width=0.23\textwidth]{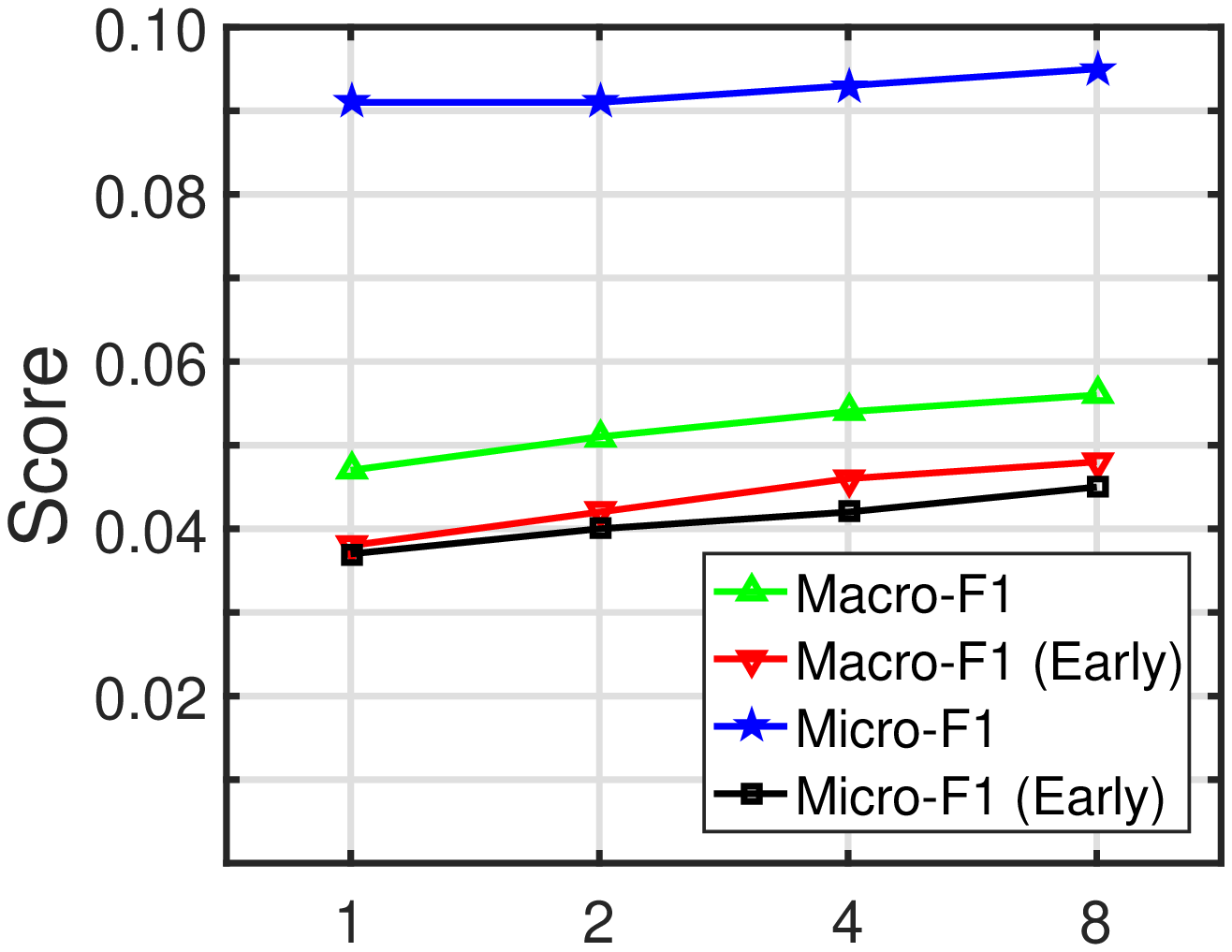}
}
\subfigure[\# dimensions $d\in\{16,32,64,128\}$]{
\includegraphics[width=0.23\textwidth]{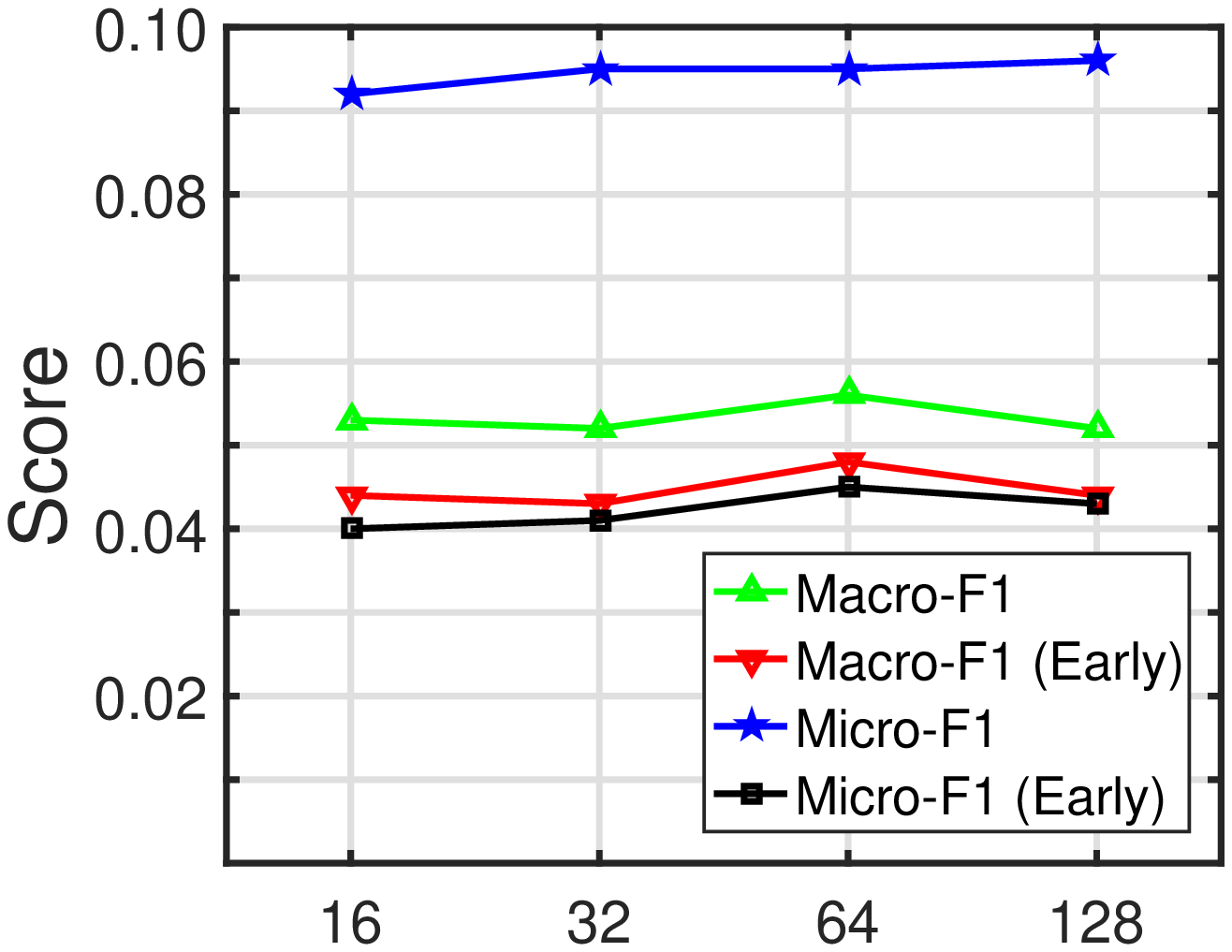}
}
\subfigure[\# window size $\textit{win}\in\{2,3,4,5\}$]{
\includegraphics[width=0.23\textwidth]{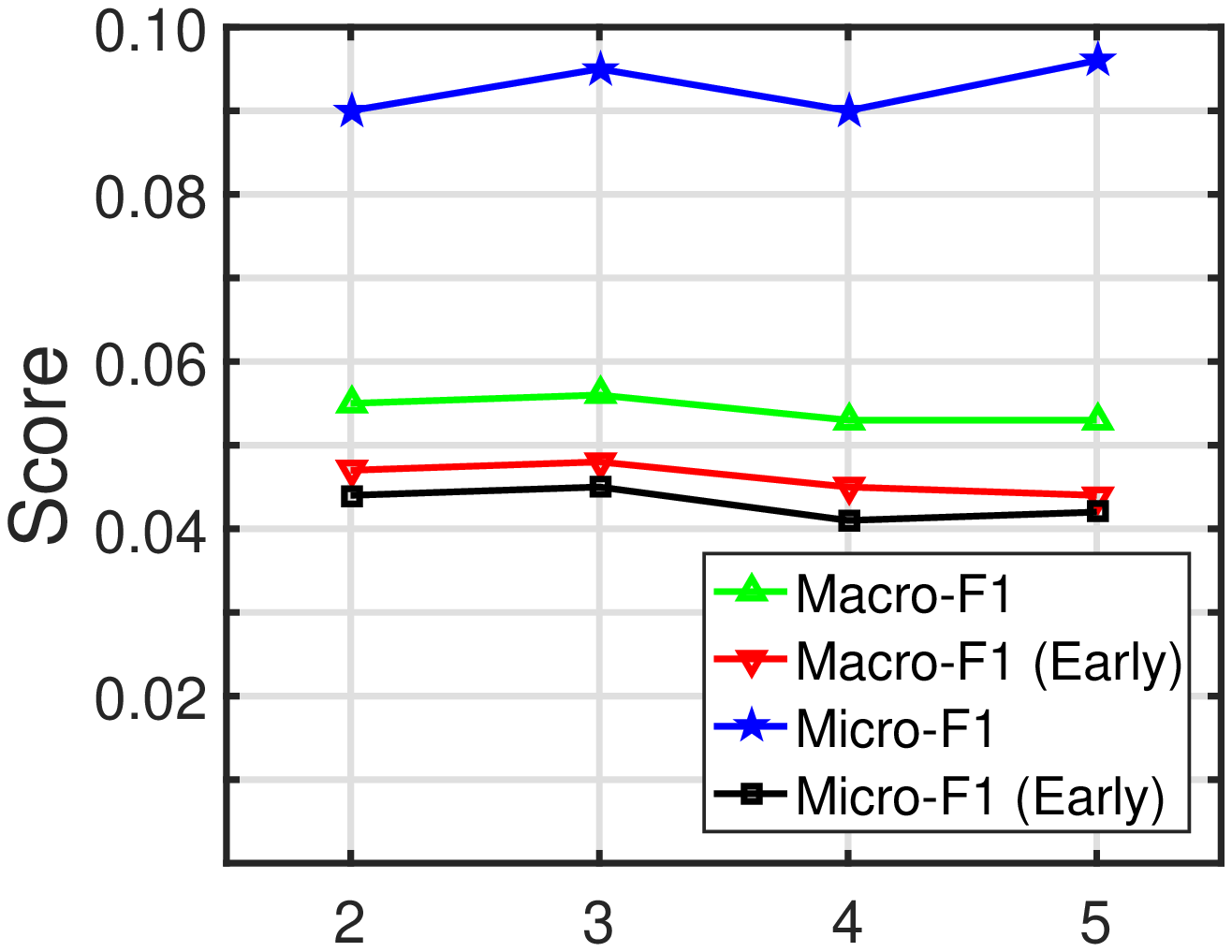}
}
\subfigure[Flag $F_{init}\in\{0,1\}$]{
\includegraphics[width=0.23\textwidth]{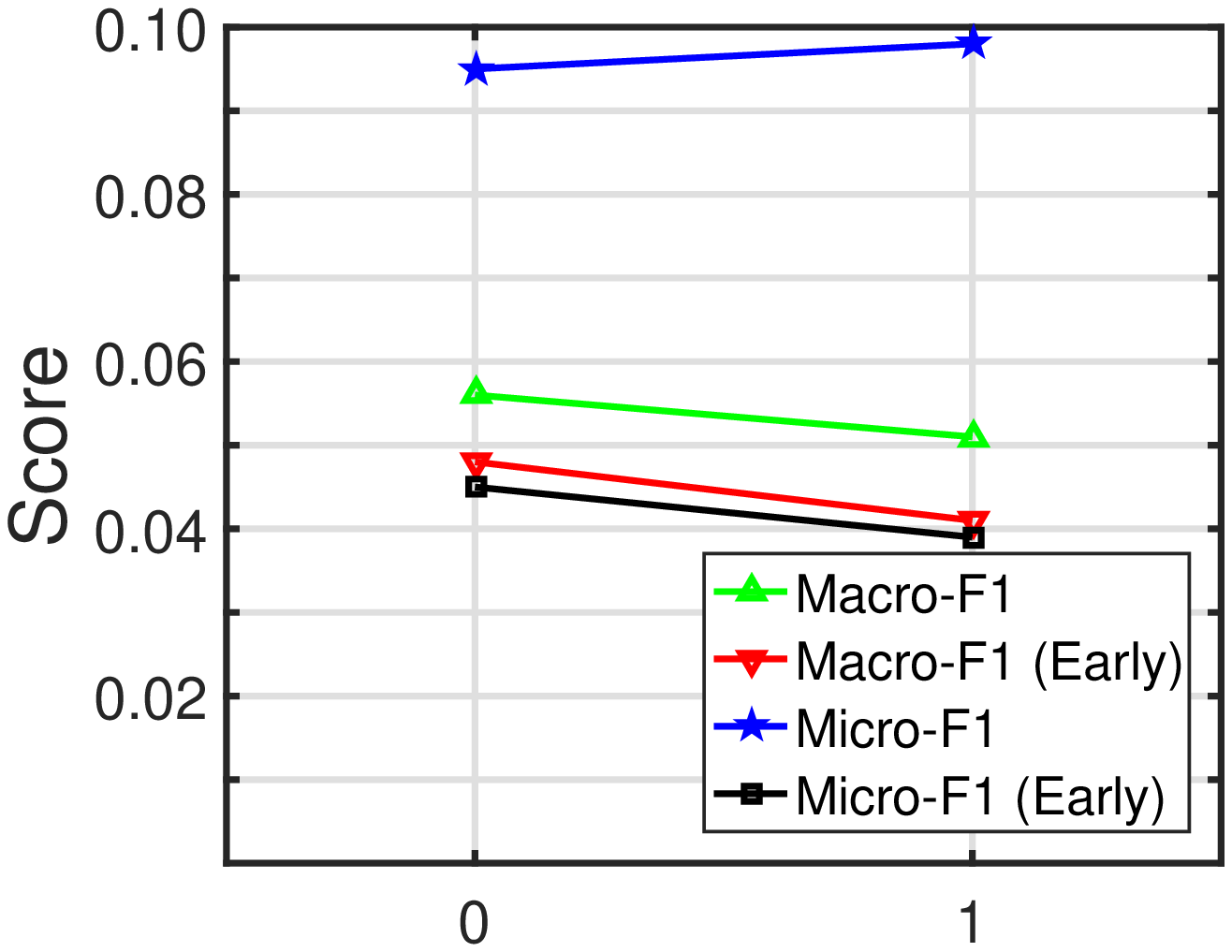}
}
\end{minipage}
\caption{Parameter sensitivity of hyperparameters in Lastfm Dataset. Macro-F1 (Early) and Micro-F1 (Early) correspond to the prediction performance at early stage.} \label{fig:ps}
\end{figure*}

Firstly, we embed the topological social network structure into real-valued user features by DeepWalk~\cite{perozzi2014deepwalk}, a widely used network representation learning algorithm. The dimension of network embeddings learned by DeepWalk is set to $32$ which is half of the dimension $d=64$ which is the representation size of our model. Secondly, we use the learned network embeddings to initialize the first $32$ dimensions of the user representations of our model and fix them during the training process without changing any other modules. In other words, a $64$-dimensional user representation is made up of a $32$-dimensional fixed network embedding learned by DeepWalk from social network structure and another $32$-dimensional randomly initialized trainable embedding. We name the modified model with Social Network considered as NDM+SN for short. This is a simple but useful implementation and we will explore a more sophisticated model to take the social network into modeling directly in future work. Fig.~\ref{fig:net2} and ~\ref{fig:net3} show the comparison between NDM and NDM+SN.

Experimental results show that NDM+SN is able to improve the performance on diffusion prediction task slightly with the help of incorporating social network structure as prior knowledge. The relative improvement of Micro-F1 is around $4\%$. The results demonstrate that our neural model is very flexible and can be easily extended to take advantage of external features. Note that these results are also consistent with those in previous subsection: The diffusion network and social network have overlapping parts but the overlapping part is relative small compared to the whole network.
\subsection{Parameter Sensitivity}
In this subsection, we will take Lastfm dataset as an illustrative example to present how hyperparameter settings affect the performance of our model. We use the best set of hyperparameter settings as our basis, \textit{i.e.} number of heads $h=8$, window size of convolutional network $\textit{win}=3$, dimension of user embeddings $d=64$ and flag of using initial user for prediction $F_{init}=0$. Then we vary each hyperparameter while keeping others fixed. Figure~\ref{fig:ps} shows the performance on diffusion prediction under different hyperparameter settings.

We can see that the performance of NDM is stable when we vary the hyperparameters within a reasonable range. NDM does not encounter serious overfitting problem when we double the dimension of embeddings $d$ to $128$. This experiment demonstrates the robustness of our model.
\subsection{Interpretability}
Admittedly, the interpretability is usually a weak point of neural network models. Compared with feature engineering methods, neural-based models encode a user into a real-valued vector space and there is no explicit meaning of each dimension of user embeddings. In our proposed model, each user embedding is projected into $16$ subspaces by an $8$-head attention mechanism. Intuitively, the user embedding in each subspace represents a specific role of the user. But it is quite hard for us to link the $16$ embeddings to interpretable hand-crafted features. We will consider the alignment between user embeddings and interpretable features based on a joint model in future work.

\begin{table}
\centering
  \caption{The scale of learned projection matrices in convolutional layer measured by Frobenius norm $||\cdot||_F^2$.}
  \begin{tabular}{ccccc}
    \toprule
    Dataset &  $W_{init}^C$& $W_0^C$&  $W_1^C$& $W_2^C$\\
    \midrule
    Lastfm & 32.3& 60.0& 49.2& 49.1\\
    Memetracker & 13.3& 16.6& 13.3& 13.0\\
    Irvine & 13.9& 13.9& 13.7& 13.7\\
    Twitter & 130.3 & 93.6& 91.5& 91.5\\
  \bottomrule
\end{tabular}
\label{tab:norm}
\end{table}

Fortunately, we still have some findings in the convolutional layer. Recall that $W_n^C\in \mathbb{R}^{d\times |\mathcal{U}|}$ for $n=0,1,2$ are position-specific linear projection matrices in convolutional layer and $W_{init}^C$ is the projection matrix for the initial user. All four matrices are randomly initialized before training. In a learned model, if the scale of one of these matrices is much larger than that of other ones, then the prediction vector is more likely to be dominated by the corresponding position. For example, if the scale of $W_0^C$ is much larger than that of other ones, then we can infer that the most recent infected user contributes most to the next infected user prediction.

Following the notations in Eq.~\ref{eq:softmax_init}, we set $F_{init}=1$ for all datasets in this experiment and compute the square of Frobenius norm of learned projection matrices as shown in Table~\ref{tab:norm}. We have the follow observations:

(1) For all four datasets, the scales of $W_0^C,W_1^C$ and $W_2^C$ are competitive and the scale of $W_0^C$ is always a little bit larger than that of the other two. This observation indicates that the active embeddings $act(u_j),act(u_{j-1}),act(u_{j-2})$ of all three recently infected users will contribute to the prediction of $u_{j+1}$. Also, the most recent infected user $u_j$ is the most important one among the three. This finding naturally matches our intuitions and verifies \textit{Assumption 2} proposed in method section.

(2) The scale of $W_{init}^C$ is the largest on Twitter dataset. This indicates that the initial user is very important in diffusion process on Twitter. This is partly because Twitter dataset contains the complete history of the spread of a URL and the initial user is actually the first one tweeting the URL. While in the other three datasets, the initial user is only the first one within the time window of crawled data. Note that we set hyperparameter $F_{init}=1$ only for Twitter dataset in diffusion prediction task because we find that the performances are competitive or even worse on the other three datasets if we set $F_{init}=1$. 
\section{Conclusion}
In this paper, we propose a Neural Diffusion Model (NDM) for microscopic cascade modeling. To go beyond the limitations of traditional cascade models based on strong assumptions and oversimplified formulas, we build our model based on two heuristic assumptions and employ deep learning techniques including convolutional neural network and attention mechanism to implement the assumptions. Experimental results on diffusion prediction task demonstrate the effectiveness and robustness of our proposed model. In addition, NDM greatly outperforms baseline methods on diffusion prediction \textit{at early stage}, which shows the applicability and feasibility of NDM for real-world applications.

For future works, we will consider linking neural-based model with hand-crafted features and statistics to improve the interpretability of learned models. An intelligible model is always welcome and can help us better understand the motivations and behaviors of users in a diffusion process.

The incorporation of extra information for cascade modeling is also an intriguing direction. For example, the timestamp information and the description of information items can be used for more accurate cascade modeling.

\ifCLASSOPTIONcompsoc
  \section*{Acknowledgments}
\else
  \section*{Acknowledgment}
\fi

This work was supported by the 973 Program (No. 2014CB340501), the Major Project of the National
Social Science Foundation of China (13\&ZD190) and the National Natural Science Foundation of China (No. 61772302). This work is also part of the NExT++ project, supported by the National Research Foundation, Prime Minister’s Office, Singapore under its IRC@Singapore Funding Initiative.

\ifCLASSOPTIONcaptionsoff
  \newpage
\fi



%
\bibliography{neuraldiffusion}
\bibliographystyle{IEEEtran}

%

\begin{IEEEbiography}[{\includegraphics[width=1in, height=1.25in, clip, keepaspectratio]{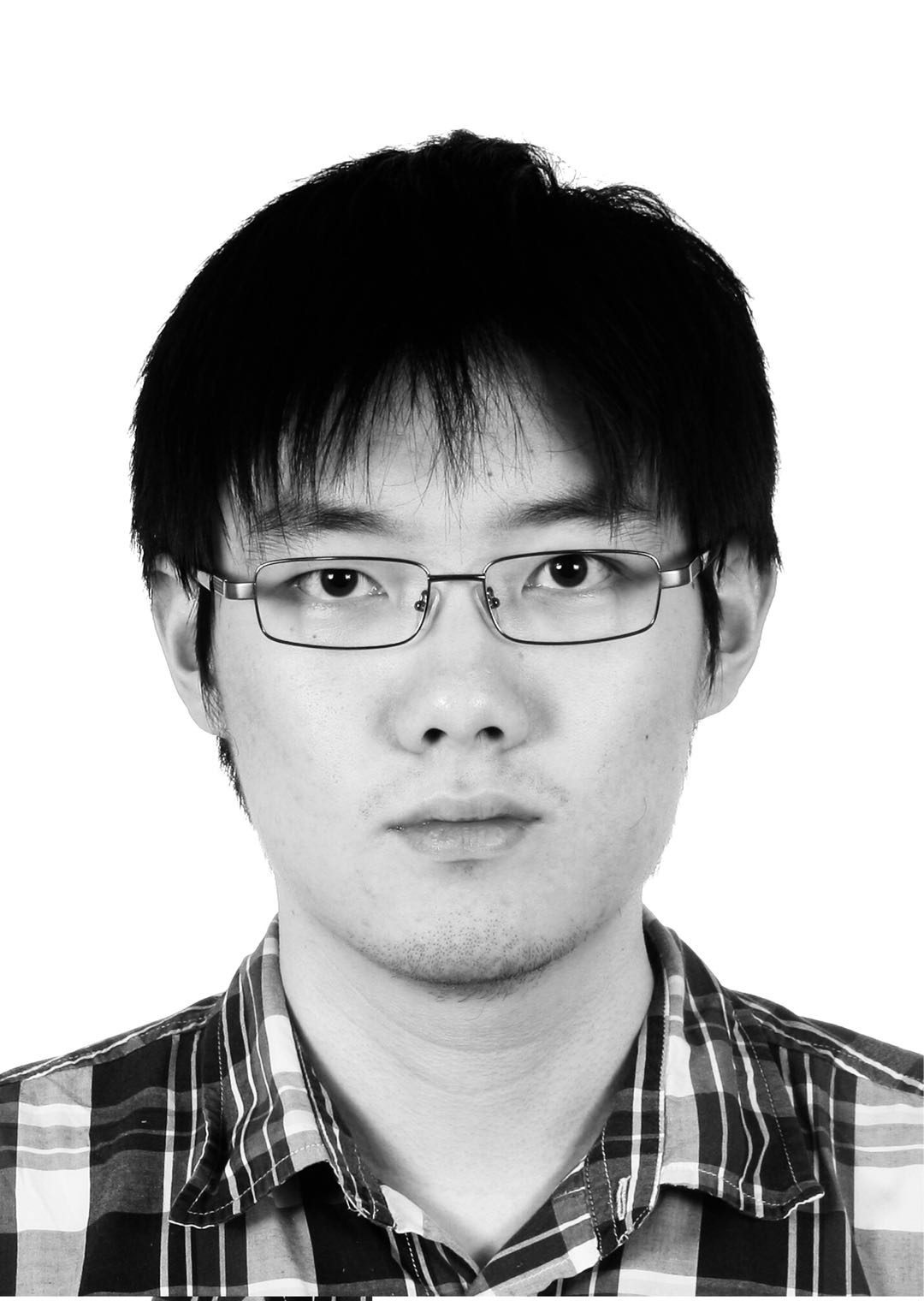}}]{Cheng Yang}
is a 4-th year PhD student of the Department of Computer Science and Technology, Tsinghua University. He got his B.E. degree from Tsinghua University in 2014. His research interests include natural language processing and network representation learning. He has published several top-level papers in international journals and conferences including ACM TOIS, IJCAI and AAAI. 
\end{IEEEbiography}

\begin{IEEEbiography}[{\includegraphics[width=1in, height=1.25in, clip, keepaspectratio]{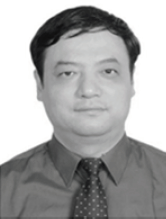}}]{Maosong Sun}
is a professor of the Department of Computer Science and Technology, Tsinghua University. He got his BEng degree in 1986 and MEng degree in 1988 from Department of Computer Science and Technology, Tsinghua University, and got his Ph.D. degree in 2004 from Department of Chinese, Translation, and Linguistics, City University of Hong Kong. His research interests include natural language processing, Chinese computing, Web intelligence, and computational social sciences. He has published over 150 papers in academic journals and international conferences in the above fields. He serves as a vice president of the Chinese Information Processing Society, the council member of China Computer Federation, the director of Massive Online Education Research Center of Tsinghua University, and the Editor-in-Chief of the Journal of Chinese Information Processing.
\end{IEEEbiography}

\begin{IEEEbiography}[{\includegraphics[width=1in, height=1.25in, clip, keepaspectratio]{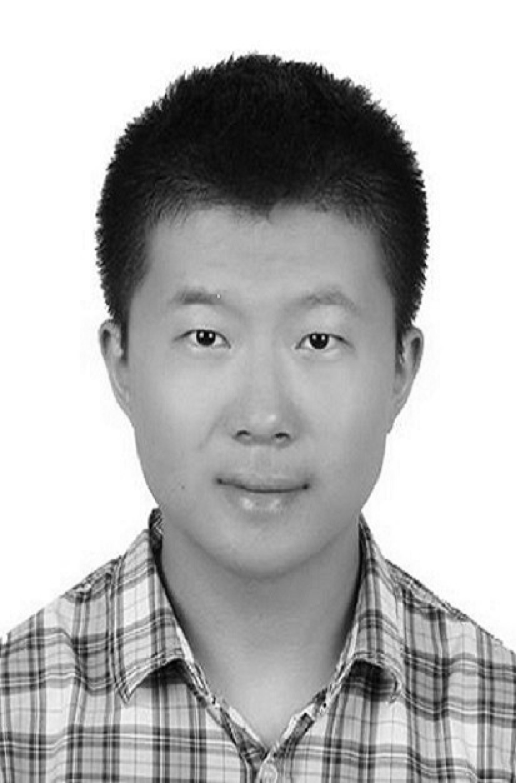}}]{Haoran Liu} is a 4-th year undergraduate student of the Department of Electric Engineering, Tsinghua University. His research interests include network representation learning and machine learning.

\end{IEEEbiography}

\begin{IEEEbiography}[{\includegraphics[width=1in, height=1.25in, clip, keepaspectratio]{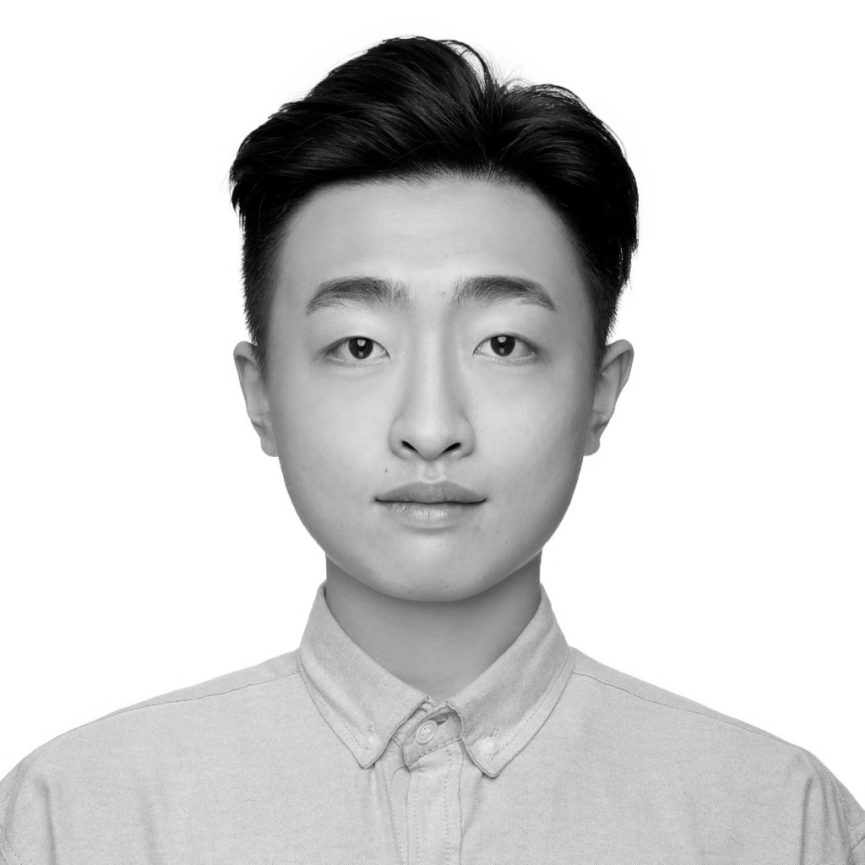}}]{Shiyi Han} is a 1-st year master student in Computer Science department at Brown University. He got his B.E. degree from Beihang University in 2018. His research interests include natural language processing and machine learning.

\end{IEEEbiography}

\begin{IEEEbiography}[{\includegraphics[width=1in, height=1.25in, clip, keepaspectratio]{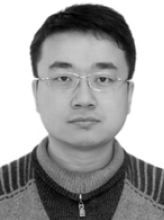}}]{Zhiyuan Liu}
is an associate professor of the Department of Computer Science and Technology, Tsinghua University. He got his BEng degree in 2006 and his Ph.D. in 2011 from the Department of Computer Science and Technology, Tsinghua University. His research interests are natural language processing and social computation. He has published over 40 papers in international journals and conferences including ACM Transactions, IJCAI, AAAI, ACL and EMNLP.
\end{IEEEbiography}

\begin{IEEEbiography}[{\includegraphics[width=1in, height=1.25in, clip, keepaspectratio]{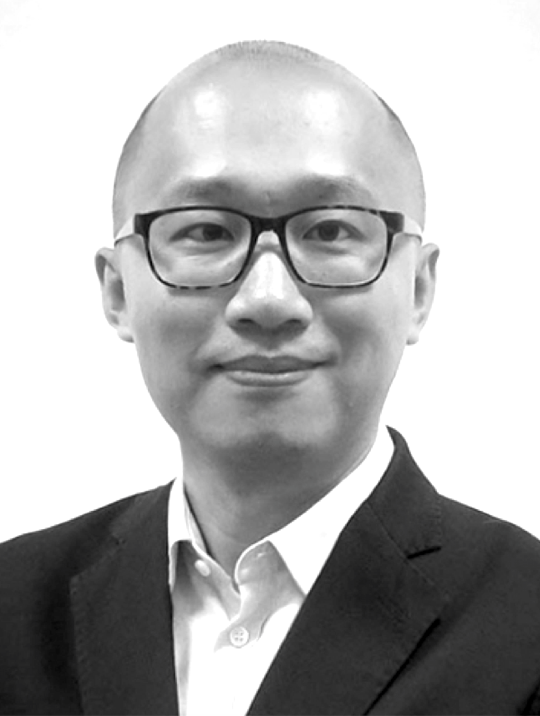}}]{Huanbo Luan} is the deputy director of NExT++ Research Center at both Tsinghua University and National University of Singapore. He received his B.S. degree in computer science from Shandong University in 2003 and Ph.D degree in computer science from Institute of Computing Technology, Chinese Academy of Sciences in 2008. His research interests include natural language processing, multimedia information retrieval, social media and big data analysis. 
\end{IEEEbiography}




\end{document}